 \documentclass[final,3p,times]{elsarticle}

\usepackage{amsmath}
\usepackage{amssymb}
\usepackage{subfigure}
\usepackage[colorlinks = true,
            linkcolor = blue,
            urlcolor  = blue,
            citecolor = red
             ]{hyperref}
\usepackage{tikz}
\usepackage{algorithm}
\usepackage{algpseudocode}
\newcommand{\tikzcircle}[2][red,fill=red]{\tikz[baseline=-0.5ex]\draw[#1,radius=#2] (0,0) circle ;}%

\graphicspath{}

\journal{Journal of Sound and Vibration}

\begin{document}

\begin{frontmatter}



\title{Improving the Efficiency of DAMAS for Sound Source Localization \\ via Wavelet Compression Computational Grid}

\author[SJTU-SAA]{Wei MA\corref{cor1}}
\ead{mawei@sjtu.edu.cn}
\cortext[cor1]{Corresponding author}
\author[KeyGo]{Xun LIU}
\address[SJTU-SAA]{School of Aeronautics and Astronautics, Shanghai Jiao Tong University, Shanghai, P. R. China}
\address[KeyGo]{Shanghai KeyGo Technology Company Limited, Shanghai, P. R. China}

\begin{abstract}
Phased microphone arrays are used widely in the applications for acoustic source localization.
Deconvolution approaches such as DAMAS successfully overcome the spatial resolution limit of the conventional delay-and-sum (DAS) beamforming method.
However deconvolution approaches require high computational effort compared to conventional DAS beamforming method.
This paper presents a novel method that serves to improve the efficiency of DAMAS via wavelet compression computational grid rather than via optimizing DAMAS algorithm.
In this method, the efficiency of DAMAS increases with compression ratio.
%
This method can thus save lots of run time in industrial applications for sound source localization, particularly when sound sources are just located in a small extent compared with scanning plane and a band of angular frequency needs to be calculated.
In addition, this method largely retains the spatial resolution of DAMAS on original computational grid, although with a minor deficiency that the occurrence probability of aliasing increasing slightly for complicated sound source.
%
%
%
\end{abstract}


\end{frontmatter}
\section{Introduction}
Nowadays the application of phased microphone arrays has become increasingly popular in the context of acoustic testing for sound source localization \cite{Michel-2006}.
The conventional delay-and-sum (DAS) beamforming algorithm computes a beamforming map of sound pressure contribution from the array microphone out signals \cite{Johnson-1993}.
The main advantages of beamforming are its simplicity and robustness.
The main disadvantage of beamforming is the poor spatial resolution particularly at low frequencies.
Additionally beamforming map is also suffers from the appearance of ghost sources due to side-lobe effects \cite{Hald-2004, Malgoezar-2016}.

To overcome these disadvantages of beamforming, deconvolution approaches have been developed.
In this kind of approaches, a source distribution is extracted from a beamforming source map by iteratively deconvolving the map.
%
The breakthrough for deconvolution approaches is DAMAS introduced by \citet{Brooks-2006} with the general assumption that the sources are not correlated among themselves, because DAMAS let deconvolution become well established in the acoustic field for ameliorating significantly the spatial resolution of beamforming map and removing thoroughly the side-lobe effects.
Unfortunately, a significant disadvantage of deconvolution is a relatively high computational effort is required compared to the conventional beamforming.

%

Reducing computational effort of deconvolution is a persistent pursuit in acoustic array measurements \cite{Bai-2013}. With less computational effort of deconvolution, it is possible to increase the dimensions of scanning map, to increase number of iterations thus improving spatial resolution, to calculate more frequency bands thus obtaining a better spectral accuracy, or to simply reduce the computational run time and thus improve the ability of real-time display and online analysis.

In order to reduce the computational effort of deconvolution, one strategy is finding more efficient deconvolution algorithms. Most of the efforts in the literature were put into this strategy.
Because spectral procedures can reduce the computational effort significantly, spectral procedures are applying in original deconvolution algorithms and then the corresponding Fourier-based algorithms are constructed. Such as DAMAS2 \cite{Dougherty-2005} and FFT-NNLS \cite{Ehrenfried-2007} are constructed by directly applying spectral procedure in DAMAS and NNLS , respectively.
In 2007 \citet{Ehrenfried-2007} compared several well-known iterative deconvolution algorithms for the mapping of acoustic sources.
Recently \citet{Lylloff-2015} first introduced an Fourier-based deconvolution algorithm FISTA in acoustic-array measurements, which is 30\% faster than FFT-NNLS.
Spectral procedures are employed under an assumption that PSF is shift-invariant, tantamount to assuming that the source field consists of plane waves.
In most aeroacoustic applications, the distance between the observation plane and the microphone array is not large compared with the extension of the region of interest, and thus the assumption that PSF is shift-invariant is invalid.
%
And thus other methods that could reduce the computation effort of the original deconvolution algorithms are urgently needed.

The computational run time of DAMAS is $O(S^2)$ for a computational grid with $S$ grid points.
%
%
Inspired by this, an alternative strategy to reduce the computational effort of DAMAS is using compression computational grid that only contains the significant grid points and does not contain the redundant grid points, if DAMAS could extract substantially similar source distribution on wavelet compression computational grid compared with that on original computational grid.
It's rational that computational grid should be compressed according to the beamformer map, which contains the information of sound sources and is already obtained before executing DAMAS.
One of the most famous compression methods is wavelet compression.
It is used naturally in adaptive wavelet collocation method for solving a large class of partial differential equations \cite{Donoho-1992, Harten-1994, Vasilyev-2000, Nejadmalayeri-2015}.
Recently wavelet compression has achieved a great success in the aerospace engineering industry \cite{Lichtl-2015} integrating with graphics processing units (GPUs).
%

The main purpose of this paper is to verify whether using wavelet compression computational grid could reduce the computation effort of DAMAS, meanwhile keep the reconstruction accuracy for sound source of DAMAS on original computational grid. The paper is organized as follows.
The general beamforming framework and DAMAS algorithm are reviewed in Sec. \ref{sec:beamforming} and Sec. \ref{sec:DAMAS}, respectively.
Wavelet compression spatial grid is illustrated in Section \ref{sec:wavelet_compression}.
Some application simulations are examined in Section \ref{sec:app_simulations}.
A discussion is presented in Section \ref{sec:discussion}.
Finally, conclusions are given in Section \ref{sec:conclusions}.


\section{Beamforming}\label{sec:beamforming}

\begin{figure}[htbp]
  \centering
  \includegraphics[width=0.5\textwidth]{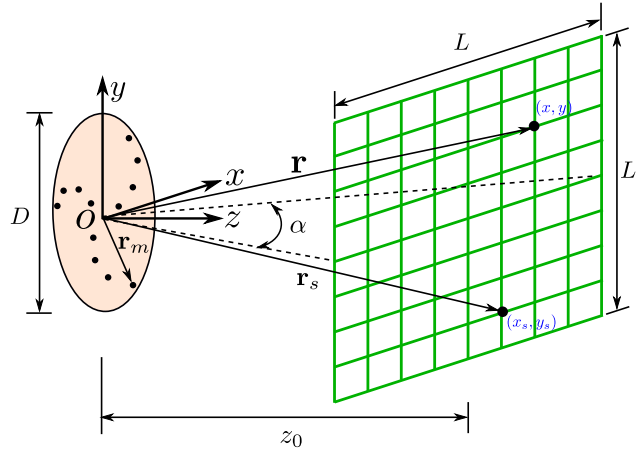}
  \caption{Sketch of setup with a microphone array. Origin of the coordinate system is placed in the centre of the microphone array.}
  \label{fig:array_illustration}
\end{figure}

Fig. \ref{fig:array_illustration} illustrates a setup with a microphone array that contains $M$ microphones and has a diameter of $D$.
Stationary noise sources are located in a $x$-$y$ plane at a distance of $z_0$ from the centre of the microphone array.
The length of the scanning plane is $L$=$2z_0\text{tan}(\alpha/2)$, where $\alpha$ is the opening angle.
The source plane is divided into $S$=$N\times N$ equidistant points.

In the traditional beamforming, cross-spectral matrix (CSM) for each test case data set is firstly calculated using simultaneously acquired data from the microphone array.
The acquired data of each microphone are divided into $I$ frames.
Each frame is then converted into frequency bins by Fast Fourier Transform (FFT).
For a given angular frequency $\omega$, CSM is averaged over $I$ frames
\begin{equation}\label{eq:CSM}
\mathbf{C}(\omega) = \overline{\mathbf{p}(\omega)\mathbf{p}(\omega)^H} = \dfrac{1}{I}\sum_{i=1}^{I}\mathbf{p}_i(\omega)\mathbf{p}_i(\omega)^H
\end{equation}
where $\mathbf{p}(\omega)=[p_1(\omega), p_2(\omega), ..., p_M(\omega)]^T$, $(\cdot)^H$ denotes conjugate transpose. For the sake of brevity, $\omega$ is omitted in the following.
Diagonal removal is usually applied to CSM, in order to improve dynamic range of the array results in poor signal-to-noise applications.

The mean-square DAS beamforming output can then be written as
\begin{equation}\label{eq:dirty_map_1}
b(\mathbf{r})=\dfrac{1}{M^2}\mathbf{v}(\mathbf{r})^H \mathbf{C} \mathbf{v}(\mathbf{r})
\end{equation}
where
\begin{equation}
\mathbf{v}(\mathbf{r})=[v_1(\mathbf{r}), v_2(\mathbf{r})), ..., v_M(\mathbf{r})]^T
\end{equation}
with the component for the $m$th microphone is
\begin{equation}
v_m(\mathbf{r})=\dfrac{|\mathbf{r}-\mathbf{r}_m|}{|\mathbf{r}|}e^{-jk|\mathbf{r}-\mathbf{r}_m|}
\end{equation}
where $|\mathbf{r}|$ is the distance from the beamformer focus position to the centre of the array, $|\mathbf{r}-\mathbf{r}_m|$ is the distance from the beamformer focus position to the $m$th microphone (see in Fig. \ref{fig:array_illustration}), and $k$ is wavenumber and $k=\omega/c_0$, where $c_0$ is speed of sound without mean flow.

\section{DAMAS}\label{sec:DAMAS}
The total sound pressure contribution at all microphones can be written as
\begin{equation}\label{eq:pGq}
\mathbf{p}=\mathbf{G}\mathbf{q}
\end{equation}
where
\begin{equation}
\mathbf{G}=
\begin{bmatrix}
g_1(\mathbf{r}_1) & g_1(\mathbf{r}_2) & \ldots & g_1(\mathbf{r}_S)  \\
g_2(\mathbf{r}_1) & g_2(\mathbf{r}_2) &  &      \\
\vdots &  & \ddots &      \\
g_M(\mathbf{r}_1) &  & & g_M(\mathbf{r}_S) \\
\end{bmatrix}
\end{equation}
is a normalized propagation matrix whose element
\begin{equation}
g_m(\mathbf{r}_s)=\dfrac{|\mathbf{r}_s|}{|\mathbf{r}_s-\mathbf{r}_m|}e^{-jk|\mathbf{r}-\mathbf{r}_m|}
\end{equation}
and $\mathbf{q}=[q_1, q_2, ..., q_S]^T$ is a vector of source amplitudes in terms of the pressure produced at the array centre.

Submitting Eq. \ref{eq:pGq} to Eq. \ref{eq:CSM}, and considering incoherent acoustic sources,
\begin{equation}
\mathbf{C}=\sum_{s=1}^{S}\overline{|q_s|^2} \cdot \mathbf{g}_{\mathbf{r}_s}\mathbf{g}_{\mathbf{r}_s}^H
\end{equation}
where $\mathbf{g}_{\mathbf{r}_s}$ is column vector of $\mathbf{G}$.
Eq. \ref{eq:dirty_map_1} can then be written as
\begin{equation}\label{eq:dirty_map_2}
b(\mathbf{r})=\sum_{s=1}^{S}\overline{|q_s|^2} \cdot \dfrac{1}{M^2}\mathbf{v}(\mathbf{r})^H
[\mathbf{g}_{\mathbf{r}_s}\mathbf{g}_{\mathbf{r}_s}^H]
\mathbf{v}(\mathbf{r})
\end{equation}
For a single unit-power point source, Eq. \ref{eq:dirty_map_2} is known as point-spread function (PSF) of the array and is defined as
\begin{equation}
\mathrm{PSF}(\mathbf{r}|\mathbf{r}_s)=\dfrac{1}{M^2}\mathbf{v}(\mathbf{r})^H
[\mathbf{g}_{\mathbf{r}_s}\mathbf{g}_{\mathbf{r}_s}^H]
\mathbf{v}(\mathbf{r})
\end{equation}
and then Eq. \ref{eq:dirty_map_2} can then be written as
\begin{equation}\label{eq:dirty_map_3}
b(\mathbf{r})=\sum_{s=1}^{S}\overline{|q_s|^2} \cdot \mathrm{PSF}(\mathbf{r}|\mathbf{r}_s)
\end{equation}

By computing $\mathrm{PSF}(\mathbf{r}|\mathbf{r}_s)$ for all combinations of $(\mathbf{r}|\mathbf{r}_s)$ in discrete grid and arranging each resulting PSF map column-wise in a matrix $\mathbf{A}$, Eq. \ref{eq:dirty_map_3} could reformulate in matrix notation as
\begin{equation}\label{eq:Axb}
\mathbf{A x=b}
\end{equation}
where $\mathbf{b}$ contains the beamformer map, and $\mathbf{x}$=$[\overline{|q_1|^2}, \overline{|q_2|^2}, ..., \overline{|q_S|^2}]^T$ is the source distribution of power descriptors.
Eq. \ref{eq:Axb} is a system of linear equations.
Notice that  $\mathbf{A} \in \mathbb{C}^{S \times S}$, $\mathbf{x} \in \mathbb{C}^{S \times 1}$, $\mathbf{b} \in \mathbb{C}^{S \times 1}$.
The deconvolution task is to find a source distribution $\mathbf{x}$ for a give dirty map $\mathbf{b}$ and know matrix $\mathbf{A}$.
The constraint is that each component of the vector $\mathbf{x}$ is larger or equal to zero.
In most of the applications the matrix $\mathbf{A}$ is singular, and $\mathbf{b}$ is in the range of $\mathbf{A}$, this means there are very large number of solutions of $\mathbf{x}$ that fulfil Eq. \ref{eq:Axb}.

The original DAMAS algorithm \cite{Brooks-2006} is an iterative algebraic deconvolution method.
In this algorithm, the source distribution is calculated by the solution of Eq. \ref{eq:Axb} using a Gauss-Seidel-type relaxation.
In each step the constraint is applied that the source strength remains positive.
The iteration step from solution $\mathbf{x}^{(n)}$ to  $\mathbf{x}^{(n+1)}$ is given by the successive application of the scheme
\begin{equation}
r_i^{(n)}=\sum_{j=1}^{i-1}A_{ij}x_j^{(n+1)}+\sum_{j=i}^lA_{ij}x_j^{(n)}-b_i
\end{equation}
and
\begin{equation}
x_i^{(n+1)}=\mathrm{max}\left( x_i^{(n)}-\dfrac{r_i^{(n)}}{A_{ii}},0 \right)
\end{equation}
for $i$=1,...,S. The values $A_{ij}$, $x_j^{(n)}$ and $b_i$ are components of the matrix $\mathbf{A}$ and the vectors $\mathbf{x}^{(n)}$ and $\mathbf{b}$, respectively.
Typically $\mathbf{x}^0$=$\mathbf{0}$ is taken as initial solution.
The value $r_i^{(n)}$ can be considered as the residual of the $i$th component in the step $n$.

The computation grid for DAMAS needs fine enough such that individual grid points along with other grid points represent a reasonable physical distribution of sources.
In addition, too coarse grid may induce spatial aliasing from source distribution.
In order to avoid aliasing problems, \citet{Brooks-2006} recommended that $\Delta x/B\leqslant0.2$, where $\Delta x$ is spacing of grid points and $B$ is array beamwidth of 3 dB down from beam peak maximum.


\section{Wavelet Compression Computational Grid}\label{sec:wavelet_compression}
In order to introduce the wavelet compression, we firstly introduce the one-dimensional wavelet that are constructed on a map $\Omega$. For more detains we refer the reader to the review of \citet{Schneider-2010}.
 The construction is performed on an arbitrary set of interpolating points, \{$x_k^j \in \Omega$\}, which are used to form a set of nested grids
\begin{equation}
  \mathcal{G}^j=\{x_k^j \in \Omega:x_k^j=x_{2k}^{j+1}, k\in \mathcal{Z}\}
\end{equation}
where $x_k^j$ are the grid points on the $j$ level of resolution.
The restriction $x_k^j=x_{2k}^{j+1}$ guarantees that $\mathcal{G}^j \subset \mathcal{G}^{j+1}$.
A function $f(x)$ on a set of nested grids with maximum level of $J$ can be decomposed as
\begin{equation}
f(x)=\sum_{k\in\mathcal{K}^0}c_k^0\phi_k^0(x)+\sum_{j=0}^{J-1}\sum_{l\in{\mathcal{L}^{j}}}d_{l}^{j}\psi_{l}^{j}(x)
\end{equation}
where $\phi_k^j(x)$ ($k\in\mathcal{K}^j$) and $\psi_{l}^{j}(x)$ ($l\in{\mathcal{L}^{j}}$) are one-dimensional tensor product scaling functions and wavelets of different families, ${\mathcal{K}^{j}}$ and ${\mathcal{L}^{j}}$ are some index sets associated respectively with scaling functions and wavelets of level $j$, $c_k^j$ and $d_k^j$ are respectively scaling function coefficients and wavelet coefficients.
If most wavelet coefficients are small for a function, a good approximation can then be retained even after discarding a large number of wavelets with small coefficients.
Inspired by this, the function $f(x)$ can be written as the sum of two terms composed of wavelets whose amplitudes are respectively above and blow some prescribed relative threshold parameter $\epsilon$,
\begin{equation}
f(x)=f_{\geqslant}(x) + f_{<}(x)
\end{equation}
where
\begin{equation}
f_{\geqslant}(x)=\sum_{k\in\mathcal{K}^0}c_k^0\phi_k^0(x)+\sum_{j=0}^{J-1}\sum_{\substack{{l\in{\mathcal{L}^{j}}}\\ |d_{l}^{j}|\geqslant \epsilon}}d_{l}^{j}\psi_{l}^{j}(x)
\end{equation}
\begin{equation}
f_{<}(x)=\sum_{j=0}^{J-1}\sum_{\substack{{l\in{\mathcal{L}^{j}}}\\ |d_{l}^{j}|< \epsilon}}d_{l}^{j}\psi_{l}^{j}(x)
\end{equation}
Then according to reference \cite{Donoho-1992},
\begin{equation}\label{eq:wavelet_10}
|f(x)-f_{\geqslant}(x)| \leqslant C \epsilon
\end{equation}
with $C$ of order unity.
Eq. \ref{eq:wavelet_10} means that discarding $f_{<}$ yields the sparse wavelet representation $f_{\geqslant}$, with the local approximation error being of the order of $\epsilon$ at any point.
In this manner a function can be represented using a minimal number of basic functions while maintaining a prescribed accuracy for a resolution $J$ sufficiently large.
Subsequently, by only reserving the grid points associated with wavelet amplitudes not less than $\epsilon$ at each resolution, an irregular sparse computational grid of essential points $\mathcal{G}_{\geqslant}$ can be obtained
\begin{equation}
\mathcal{G}_{\geqslant}=\sum_{j=0}^{J}\mathcal{G}_{\geqslant}^{j}
\end{equation}
where
\begin{equation}\label{eq:wavelet_epsilon}
\mathcal{G}_{\geqslant}^{j}=\{x_k^j \in \mathcal{G}^j:d_k^j\geqslant \epsilon\}
\end{equation}
Note that grid points on the coarsest level are only associated with scaling functions and thus are always in $\mathcal{G}_{\geqslant}$.
In the interpolating wavelet \cite{Deslauriers-1989, Donoho-1992, Harten-1994, Wirasaet-2005}, wavelet coefficients can be fast calculated as the difference between the actual value at a point and its interpolated value using only information on the adjacent coarser grid level,
\begin{equation}
d_k^j=f(x_k^j)-\overline{f(x_k^j)}
\end{equation}
where $x_k^j\in \{\mathcal{G}^{j} \circleddash \mathcal{G}^{j-1}\}$, $\overline{f(x_k^j)}$ is the interpolating value using only information on $\mathcal{G}^{j-1}$.
The one-dimensional wavelet transform described above can be easily extended to multiple dimensions using tensor product construction.

\begin{figure}[htbp]
  \centering
  \includegraphics[width=0.4\textwidth]{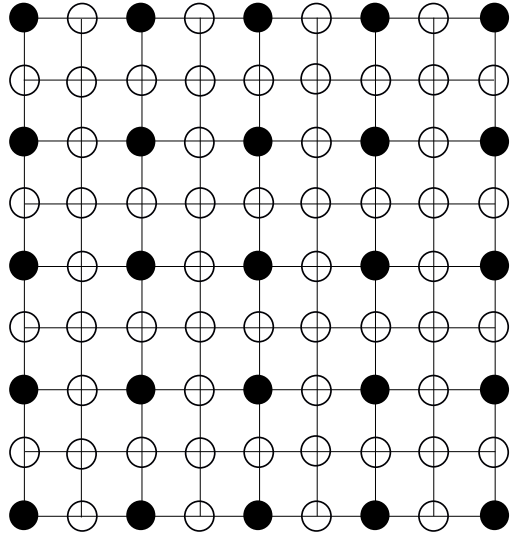}
  \caption{Sketch of wavelet compression. Solid points (\tikzcircle[black,fill=black]{3pt}), reference points;  circle points ($\bigcirc$), calibration points.}
  \label{fig:compressed_grid}
\end{figure}

In this paper the computational grid for DAMAS is compressed by interpolation wavelet compression according to the beamformer map, because beamformer map contains the information of sound sources and is already obtained before executing DAMAS.
Algorithm \ref{alg:wavelet_compression} show the processing to find the grid points that need to be included in the wavelet compression computational grid.
The original computational grid for beamformer map is considered as the finest grid.
Fig. \ref{fig:compressed_grid} sketches the compression processing at a level of resolution. $\mathbf{b}$ is interpolated on reference points (solid points in Fig. \ref{fig:compressed_grid}) to calibration points (circle points in Fig. \ref{fig:compressed_grid}). At this step, compression grid only contains the calibration points where the difference between interpolation value and original value are less than $\epsilon$.

The original linear equations (Eq. \ref{eq:Axb}) on compression grid comes
\begin{equation}\label{eq:Axb_CG}
\mathbf{\tilde{A} \tilde{x}=\tilde{b}}
\end{equation}
where $\mathbf{\tilde{A}} \in \mathbb{C}^{\tilde{S} \times \tilde{S}}$, $\mathbf{\tilde{x}} \in \mathbb{C}^{\tilde{S} \times 1}$, $\mathbf{\tilde{b}} \in \mathbb{C}^{\tilde{S} \times 1}$, $\tilde{S}$ is the number of compression computational grid points.

\begin{algorithm}[h]
\caption{Wavelet Compression Computational Grid}
\begin{algorithmic}[1]
\State Maximum level of resolution $J$ = the round down to the next integer of $\text{log}_2{N}$
\State Form a set of nested grids $\mathcal{G}^j=\{(x_k^j,y_k^j) \in \Omega_{\mathbf{b}}:x_k^j=x_{2k}^{j+1}, y_k^j=y_{2k}^{j+1}, k\in \mathcal{Z}\}$, $j$=0, ... , $J$
\For{$j$ = $J-1$:-1:0}
\State $\mathcal{G}_{\geqslant}^{j+1}$=$\varnothing$
\State Interpolate $\mathbf{b}$ on grid $\mathcal{G}^j$ to grid $\{\mathcal{G}^{j+1} \circleddash \mathcal{G}^{j}\}  $;
\For{Any grid point $(x,y)\in\{\mathcal{G}^{j+1} \circleddash \mathcal{G}^{j}\}$}
\If{$|\overline{b^{j+1}(x,y)}-b(x,y)|\geqslant \epsilon$}
\State $\mathcal{G}_{\geqslant}^{j+1}=\mathcal{G}_{\geqslant}^{j+1} \cup p$
\EndIf
\EndFor
\EndFor
\State $\mathcal{G}_{\geqslant}^{0}=\mathcal{G}^0$
\State $\mathcal{G}_{\geqslant}$=$\varnothing$
\For {$j$ = 0:1:$J$}
\State $\mathcal{G}_{\geqslant}$=$\mathcal{G}_{\geqslant} \oplus \mathcal{G}_{\geqslant}^{j}$
\EndFor
\label{code:recentEnd}
\end{algorithmic}
\label{alg:wavelet_compression}
\end{algorithm}

\clearpage
\section{Application Simulations}\label{sec:app_simulations}
In this section four application simulations are carried out on an Intel Core i5-2500K 3.30 GHz processor.
The information of these four application simulations is listed in Table \ref{tab:appl_simul}.

The planar array contains 60 microphones and has a diameter of 1.0 m.
In the geometrical setup, the observation plane is parallel to the array plane, and the region of interest is right in front of the array.
The distance between array plane and observation plane $z_0$ is 5.0 m.
The opening angle $\alpha$=60$^{\circ}$, while the off-axis angle $\phi$=0.5$\alpha$=30$^{\circ}$.
In these simulations, the frequency $f$ is 3 kHz.
This corresponds to a wavelength of 0.113 m when the speed of sound $c_0$ is 340 m/s.
The beamformer resolution limit $B$ is about 1.06 m, according to Rayleigh's criterion \cite{Bai-2013}
\begin{equation}
B=\dfrac{1.22 z_0 c_0}{\text{cos}^3(\phi)Df}
\end{equation}
The original computational grid is 50$\times$ 50 with 2500 grid points.
The parameter $\Delta x/B$ is thus 0.11, falling in the range of [0.05, 0.2] that has been recommended by \citet{Brooks-2006}.
Gaussian white noise is added with a signal-to-noise ratio of 15 dB at the microphone array.
In each simulation, the source amplitudes $q_s$ are assigned in the source distribution of power descriptors $\mathbf{x}$=$[\overline{|q_1|^2}, \overline{|q_2|^2}, ..., \overline{|q_S|^2}]^T$.
The corresponding beamformer map $\mathbf{b}$ is then obtained according to Eq. \ref{eq:Axb}.
When applying DAMAS the deconvolution algorithm, the starting guess $\mathbf{x}^0$=0, and the number of iteration is 1000.
The deconvolved maps already converge after 1000 iterations.
Compression computational grid is obtained by wavelet compression introduced in Section \ref{sec:wavelet_compression} according to beamformer map.
And then DAMAS is executed on compression grid.

\begin{table}[htbp]
\centering
\caption{Information of application simulations}
\label{tab:appl_simul}
\begin{tabular}{ccccc}
\hline\hline
Number of case         & Case 1 & Case 2 & Case 3 & Case 4 \\ \hline\hline
Arrray aperture diameter, $D$ (m) &  \multicolumn{4}{c}{1.0}  \\
Dis. between array plane and observation plane, $z_0$ (m) & \multicolumn{4}{c}{5.0} \\
Opening angle, $\alpha$ & \multicolumn{4}{c}{60$^{\circ}$} \\
Scanning length, $L=2z_0\text{tan}(\alpha/2)$ (m) & \multicolumn{4}{c}{5.77} \\
Frequency, $f$ (kHz)       & \multicolumn{4}{c}{3} \\
Beamformer resolution, $B$ (m) & \multicolumn{4}{c}{1.06} \\
Original Grid & \multicolumn{4}{c}{50$\times$50} \\
Number of original grid points, $S$ &\multicolumn{4}{c}{2500} \\
$\Delta x/B$ & \multicolumn{4}{c}{0.11} \\
\hline
Source power style & One point & Two points & Two points & DAMAS image \\
Integrated source power setting advance, $P_0$ (Pa) & 1.000 & 1.000+1.000 &1.000+0.316 & 70$\times$1.000 \\
\hline
Run time on original grid after 1000 itera., $T_1$ (s) & 102 & 102 & 102 & 102 \\
Integrated source power on original grid, $P_1$ (Pa) & 0.992 & 0.970+0.990 & 0.980+0.287 & 51.100\\
Error of integ. source power on ori. grid, $\eta$=$\dfrac{P_0-P_1}{P_0}$ & 0.8\% & 2.1\% &3.7\% & 27.0\%\\
\hline
$\epsilon$ & 0.1 & 0.3 & 0.1 & 0.1\\
Number of compr. grid points, $\tilde{S}$ & 17 & 16 & 17 & 543 \\
Compression ratio, $\sigma=S/\tilde{S}$ & 147 & 156 & 147 & 4.6 \\
Time of grid compression (s) & 0.055 & 0.055 & 0.055 & 0.055  \\
Run time on compr. grid after 1000 iterations, $T_2$ (s) & 0.429 & 0.396 & 0.424 & 17.6 \\
Efficiency increasing, $\dfrac{T_1-T_2}{T_1}$ & 99.6\% & 99.6\% & 99.6\% & 82.8\% \\
Integrated source power on compression grid, $P_2$ (Pa) & 0.980 & 1.223+0.706 & 0.983+0.278 & 49.38\\
Error of integ. source power on comp. grid, $\eta$=$\dfrac{P_0-P_2}{P_0}$ & 2.0\% & 3.5\% &4.3\% & 29.5\% \\
\hline\hline
\end{tabular}
\end{table}

\subsection{One point source}
\begin{figure}[htbp]
\centering
  \subfigure[]{
    \label{fig:beamformer_map} 
    \includegraphics[width=0.49\textwidth]{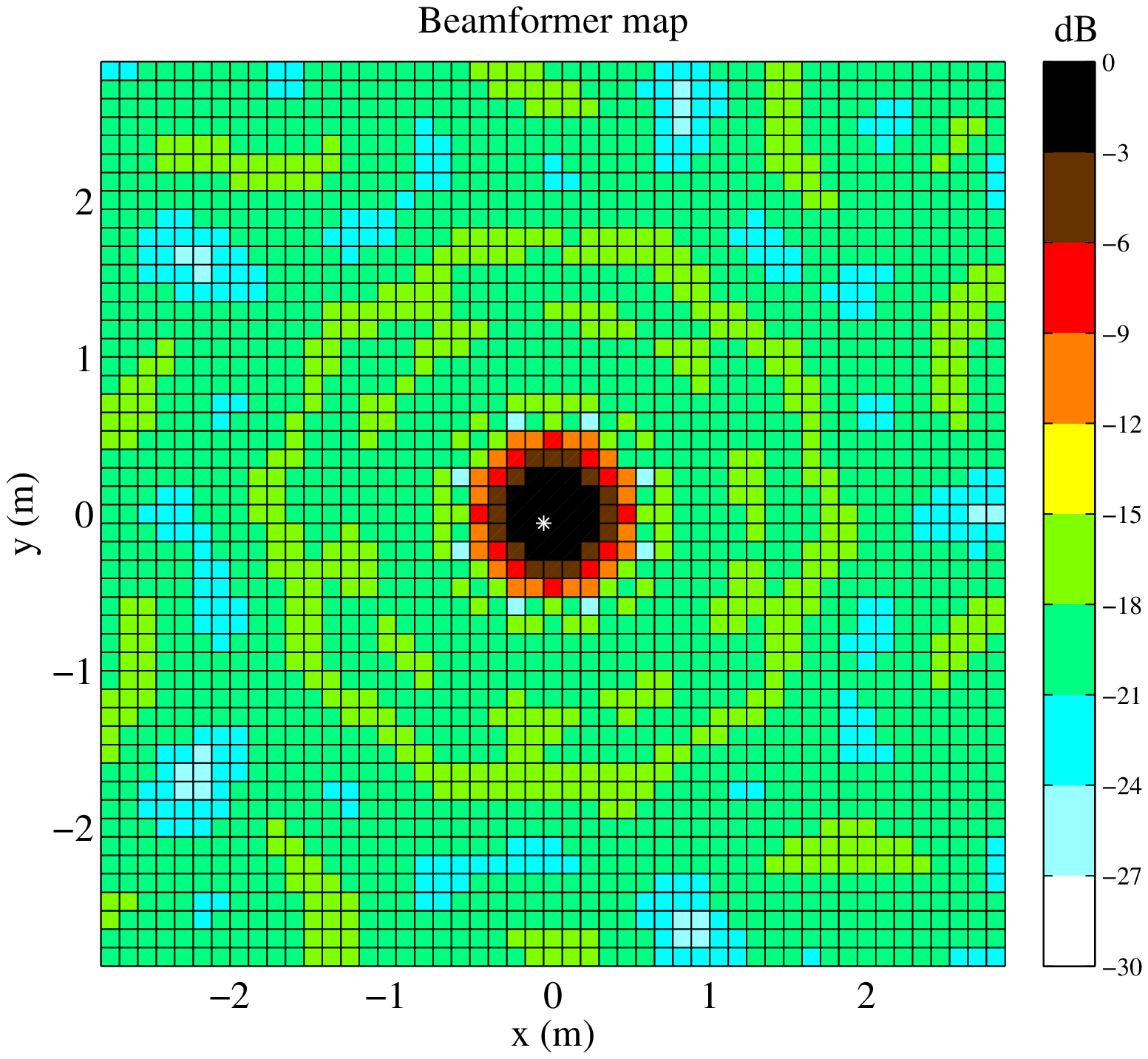}}
  \subfigure[]{
    \label{fig:DAMAS} 
    \includegraphics[width=0.49\textwidth]{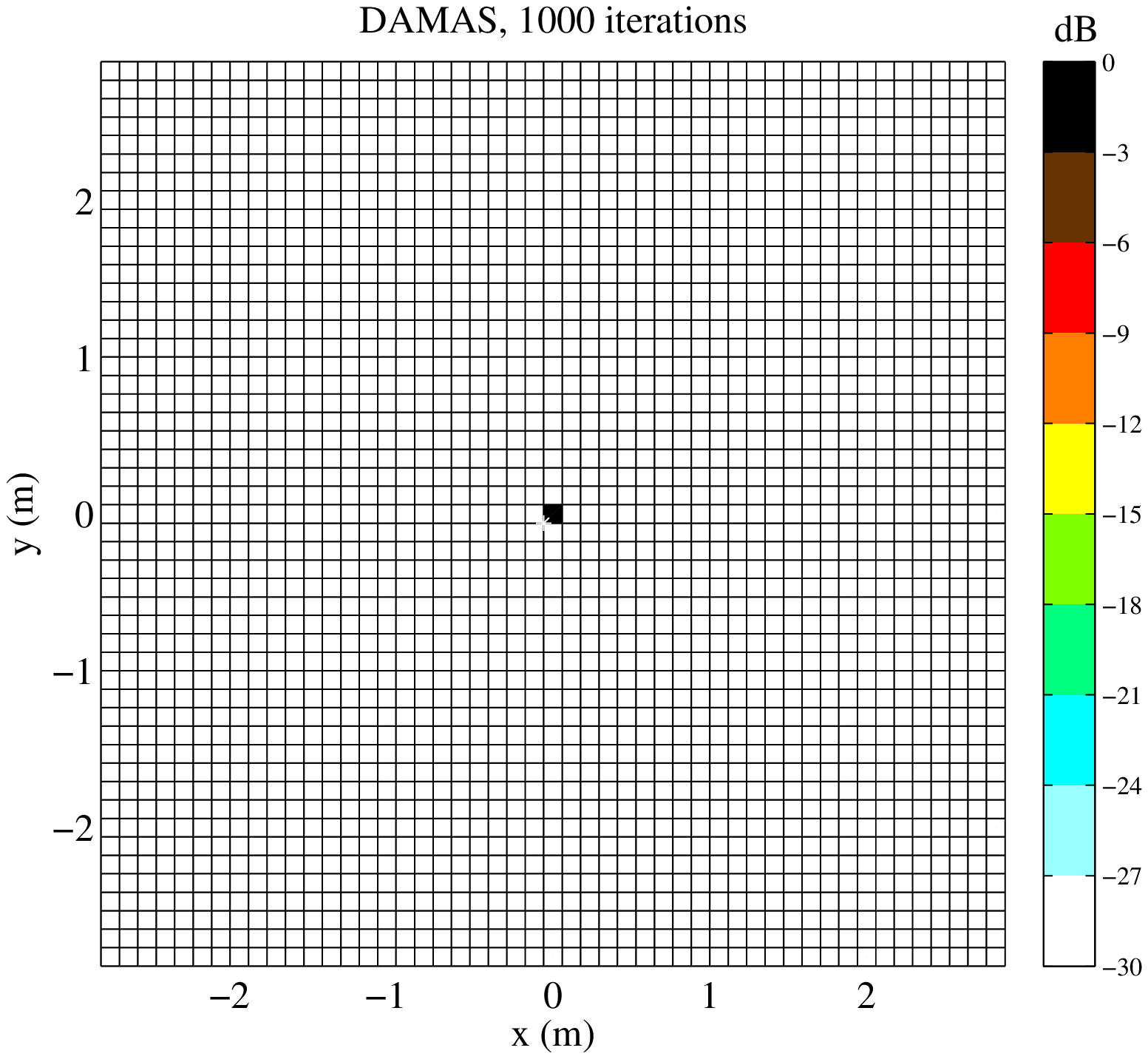}}\\
  \subfigure[]{
    \label{fig:beamformer_map} 
    \includegraphics[width=0.49\textwidth]{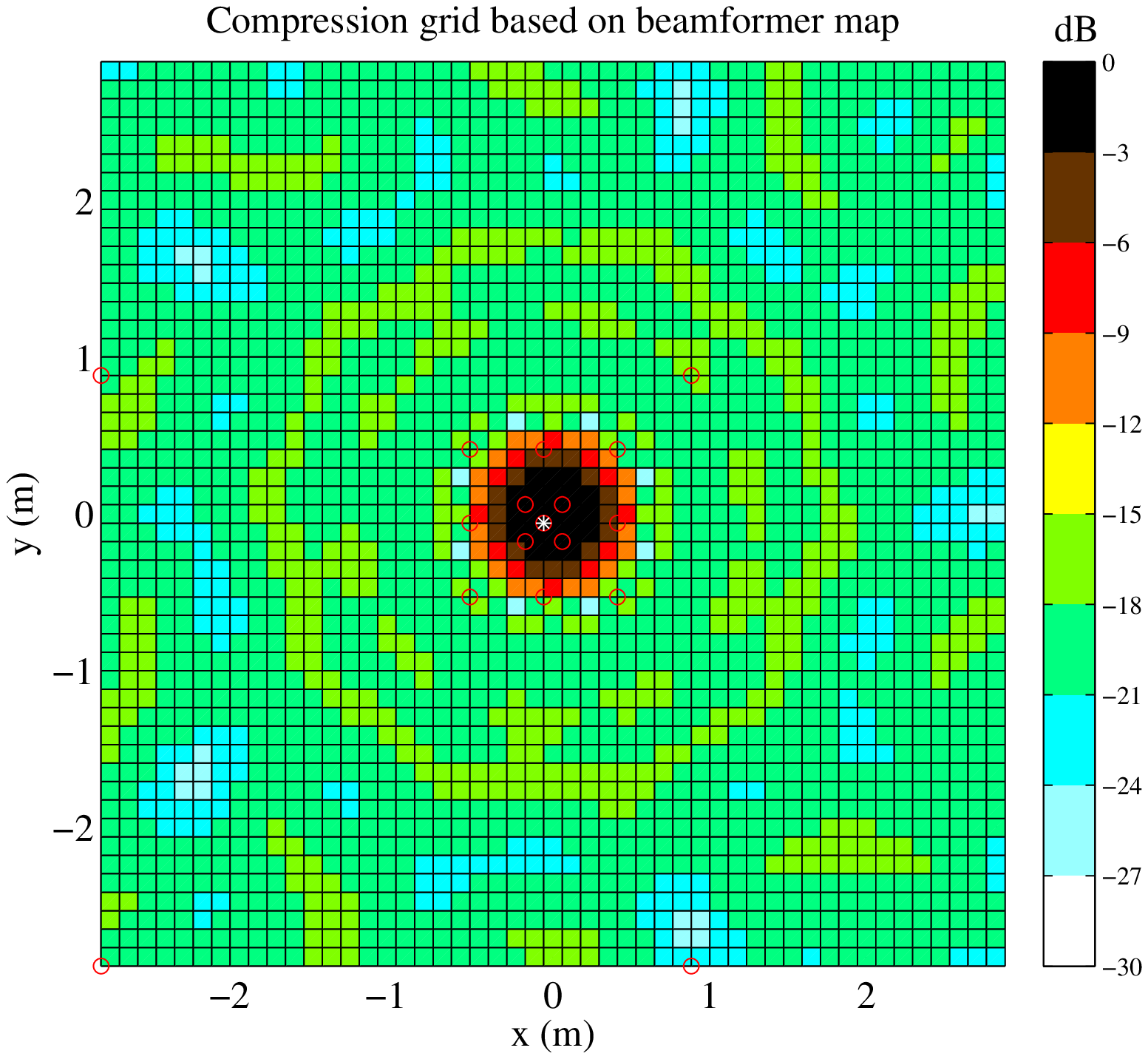}}
  \subfigure[]{
    \label{fig:DAMAS} 
    \includegraphics[width=0.49\textwidth]{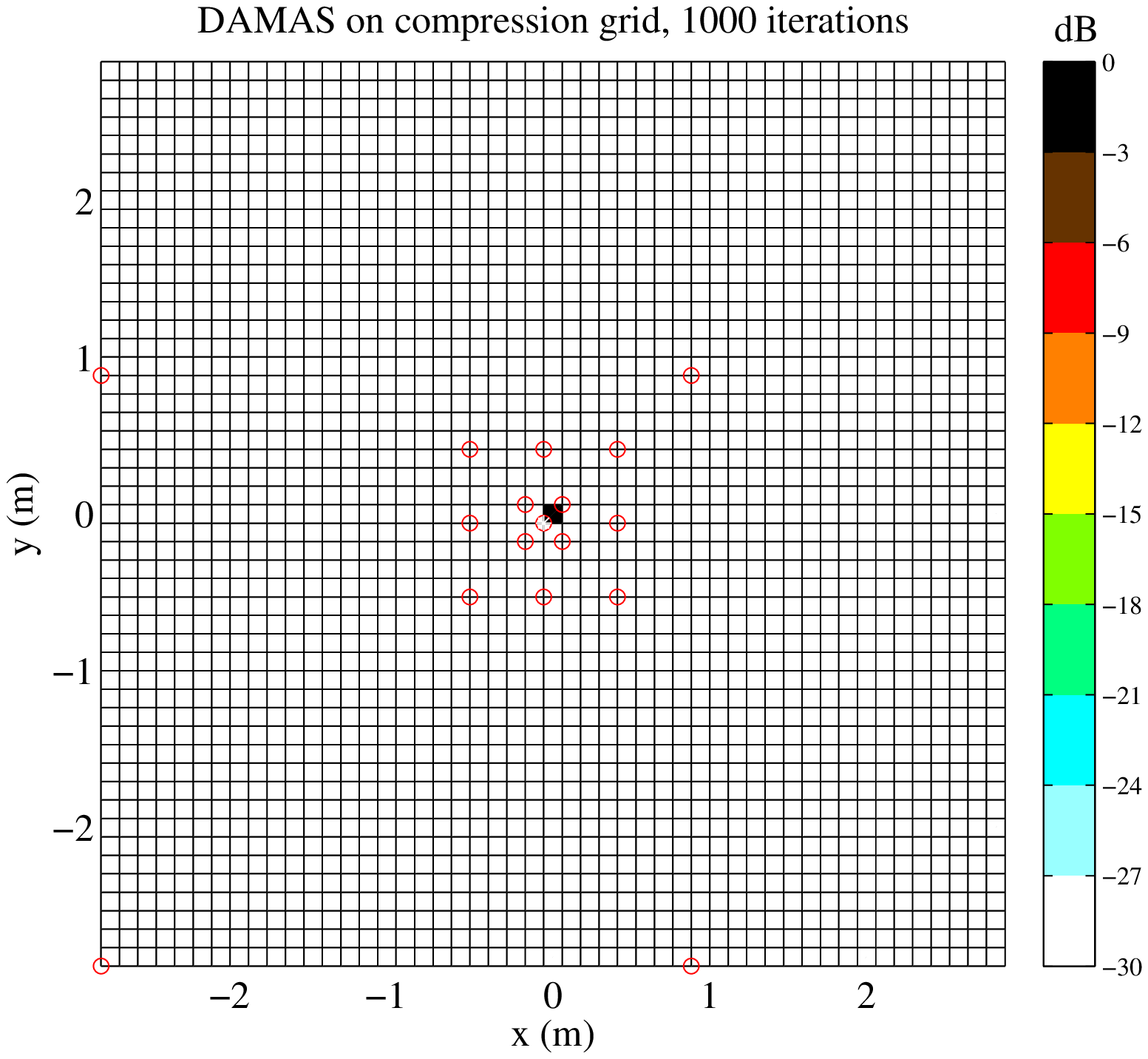}}\\
\caption[]{Synthetic one point source, $f$=3 kHz. White snow symbol, position of synthetic point source. (a) Beamformer map on original grid. (b) Deconvolved map of DAMAS on original grid. (c) Compressed grid based on beamformer map, shown by red circles. Compression ratio, 278. (d) Deconvolved map of DAMAS on compression grid.}
\label{fig:NO1}
\end{figure}

In first simulation, a single synthetic point source is placed at grid point (25, 25) in the centre of the plane.
The source amplitudes $q_s$ are defined to 1.000 Pa at the source point and 0 at all other grid points.
The corresponding beamformer map on original grid is shown in Fig. \ref{fig:NO1}a.
The result after applying DAMAS algorithm on original grid is shown in Fig. \ref{fig:NO1}b.
The source is well resolved by DAMAS on original grid with a very small error of source power of 0.8\%. This small error of source power is due to noise signal added in the simulation.

The compression grid based on beamformer map is shown in Fig. \ref{fig:NO1}c with red circle points.
The compression ratio is about 147.
The source point is reserved in the wavelet compression grid.
A lot of grid points with zero source power are discarded as redundant points.
On compression grid, the result after applying DAMAS is shown in Fig. \ref{fig:NO1}d.
The source position is well resolved by DAMAS on this compression grid.
The error of source power on compression grid is only about 2.0\%.
From the run time listed in Table \ref{tab:appl_simul}, DAMAS takes 102 ms on original grid while only 0.429 ms on compression grid per iteration.
DAMAS is thus about 99.6\% faster on compression grid.
The time of compression grid is only 55 ms, and is only half of the run time of DAMAS on original grid for 1 iteration.
The efficiency of DAMAS has been improved significantly through compression grid.

This simulation verifies that when grid points with zero source power are discarded as redundant points in wavelet compression grid, the efficiency of DAMAS has indeed been improved through wavelet compression grid.
Meanwhile DAMAS on compression grid keeps the spatial resolution of DAMAS on original grid in this simulation with simple one point source.

\subsection{Two points source with equal source power}
\begin{figure}[htbp]
\centering
  \subfigure[]{
    \label{fig:beamformer_map} 
    \includegraphics[width=0.49\textwidth]{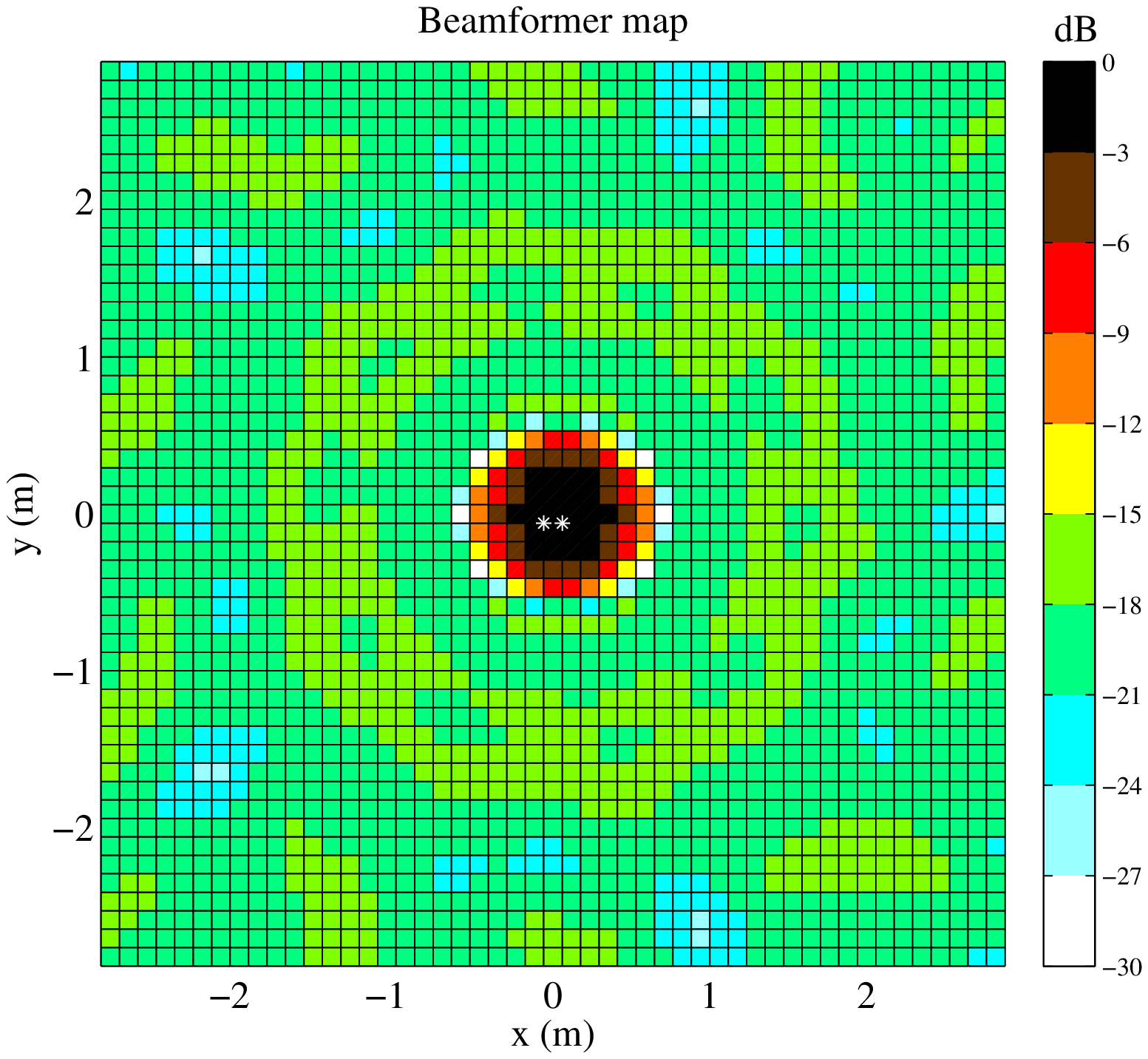}}
  \subfigure[]{
    \label{fig:DAMAS} 
    \includegraphics[width=0.49\textwidth]{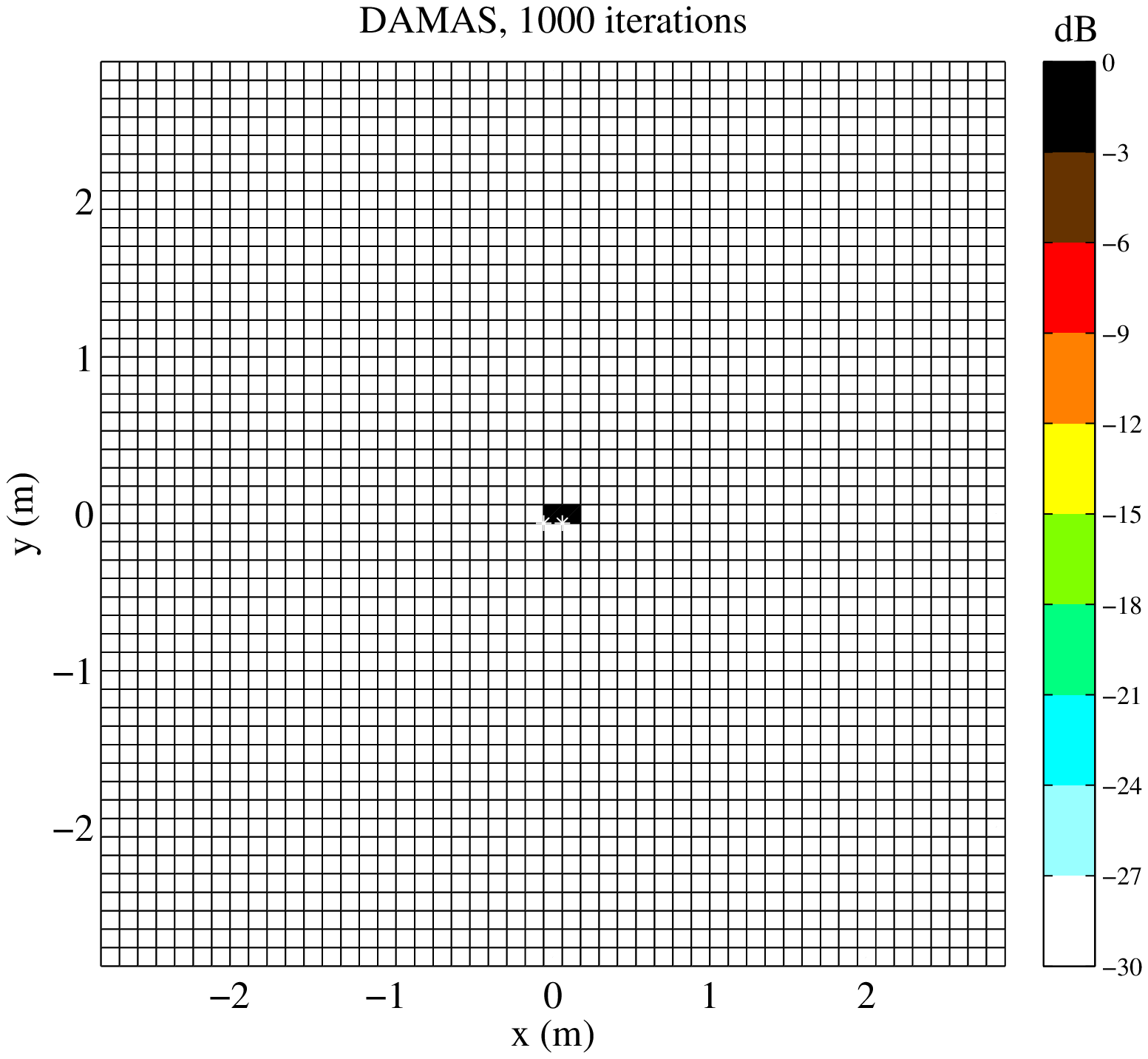}}\\
  \subfigure[]{
    \label{fig:beamformer_map} 
    \includegraphics[width=0.49\textwidth]{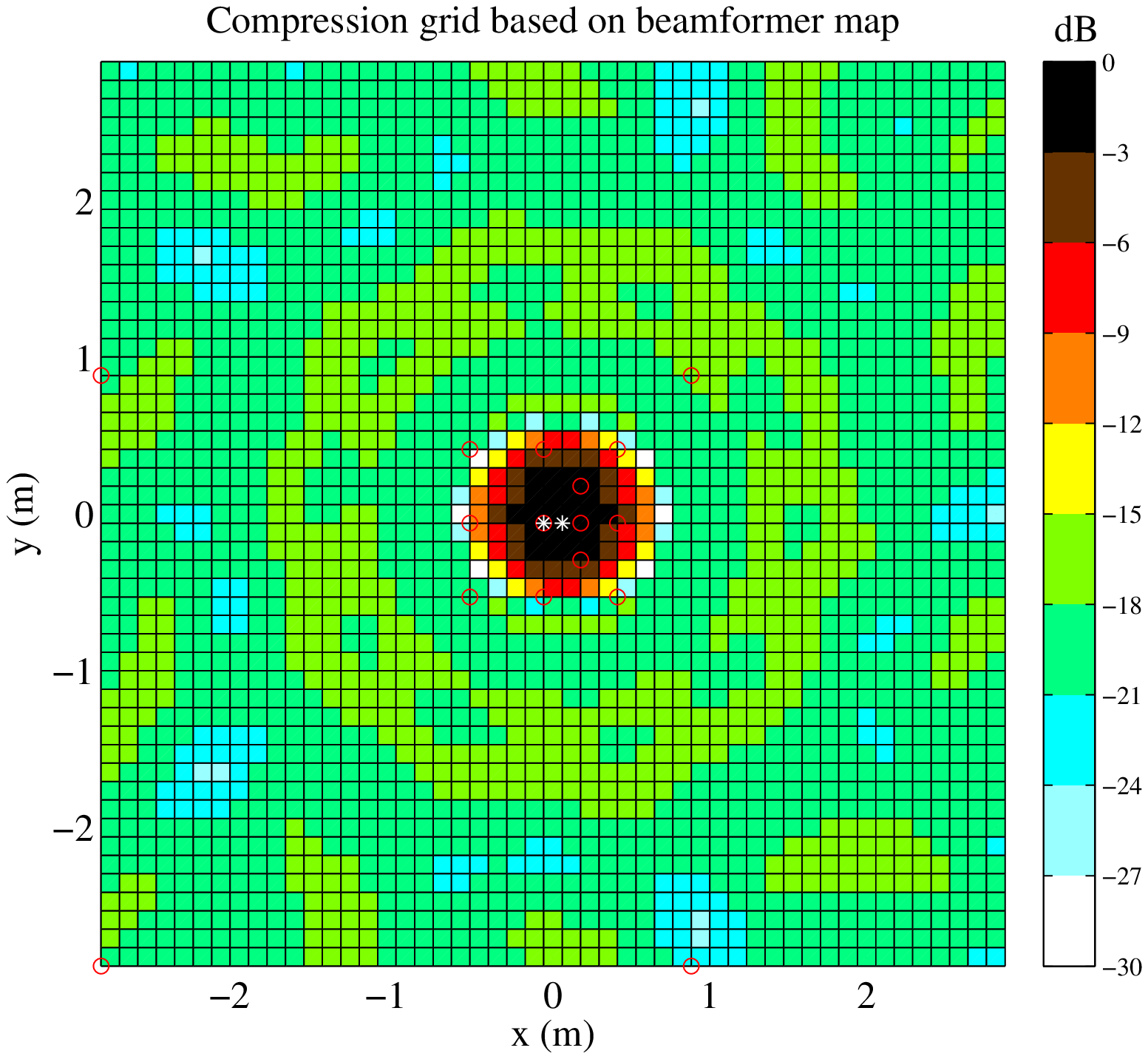}}
  \subfigure[]{
    \label{fig:DAMAS} 
    \includegraphics[width=0.49\textwidth]{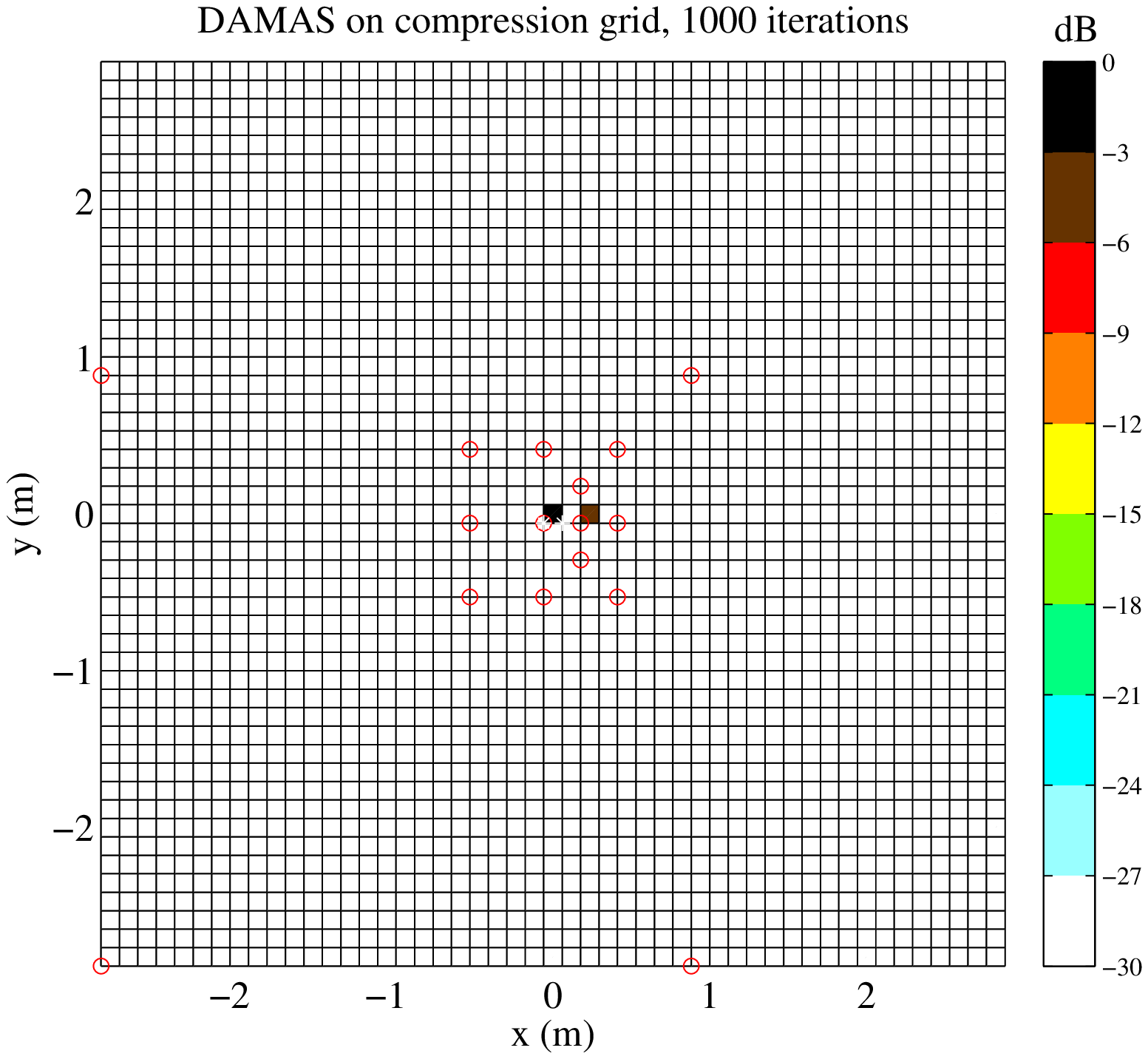}}\\
\caption[]{Synthetic two points source, $f$=3 kHz. White snow symbols, positions of synthetic point sources. (a) Beamformer map on original grid. (b) Deconvolved map of DAMAS on original grid. (c) Compression grid based on beamformer map, shown by red circles. (d) Deconvolved map of DAMAS on compression grid.}
\label{fig:NO2}
\end{figure}

In second simulation, synthetic two points source is placed at grid points (25, 25) and (26, 25), i.e. another source point is placed at the point just adjacent to the source point in the first simulation.
The source amplitudes $q_s$ are defined to 1.0 Pa at these two source points, and 0 at all other grid points.
The corresponding beamformer map on original grid is shown in Fig. \ref{fig:NO2}a.
The result after applying DAMAS algorithm on original grid is shown in Fig. \ref{fig:NO2}b.
Two source points are well resolved by DAMAS on original grid.

The compression grid based on beamformer map is shown in Fig. \ref{fig:NO2}c with red circle points.
The compression ratio is about 1562.
The source point at (25, 25) is reserved in the wavelet compression grid, while the source point at (26, 25) is discarded as a redundant point.
Meanwhile many grid points with zero source power are discarded as redundant points.
On compression grid, the result after applying DAMAS is shown in Fig. \ref{fig:NO2}d.
The resolved source power is 1.223 Pa at point (25, 25), 0.706 Pa at another adjoining point to point (27, 25), and 0 at other points.
The source power at the second point (26, 25) is distributed on two adjoining points.
The integrated source power is also well resolved with a very small error by DAMAS on the compression grid.
From the run time listed in Table \ref{tab:appl_simul}, the efficiency of DAMAS at this simulation has also been improved significantly through compression grid as that in the first simulation.

This simulation verifies that when grid points with non-zero source power are discarded as redundant points in wavelet compression grid, the efficiency of DAMAS has indeed been improved through wavelet compression grid.
At the same time, DAMAS on compression grid could extract substantially similar source distribution through wavelet compression grid compared with that on original grid.
%

\subsection{Two points source with different source powers of 10 dB}
\begin{figure}[htbp]
\centering
  \subfigure[]{
    \label{fig:beamformer_map} 
    \includegraphics[width=0.49\textwidth]{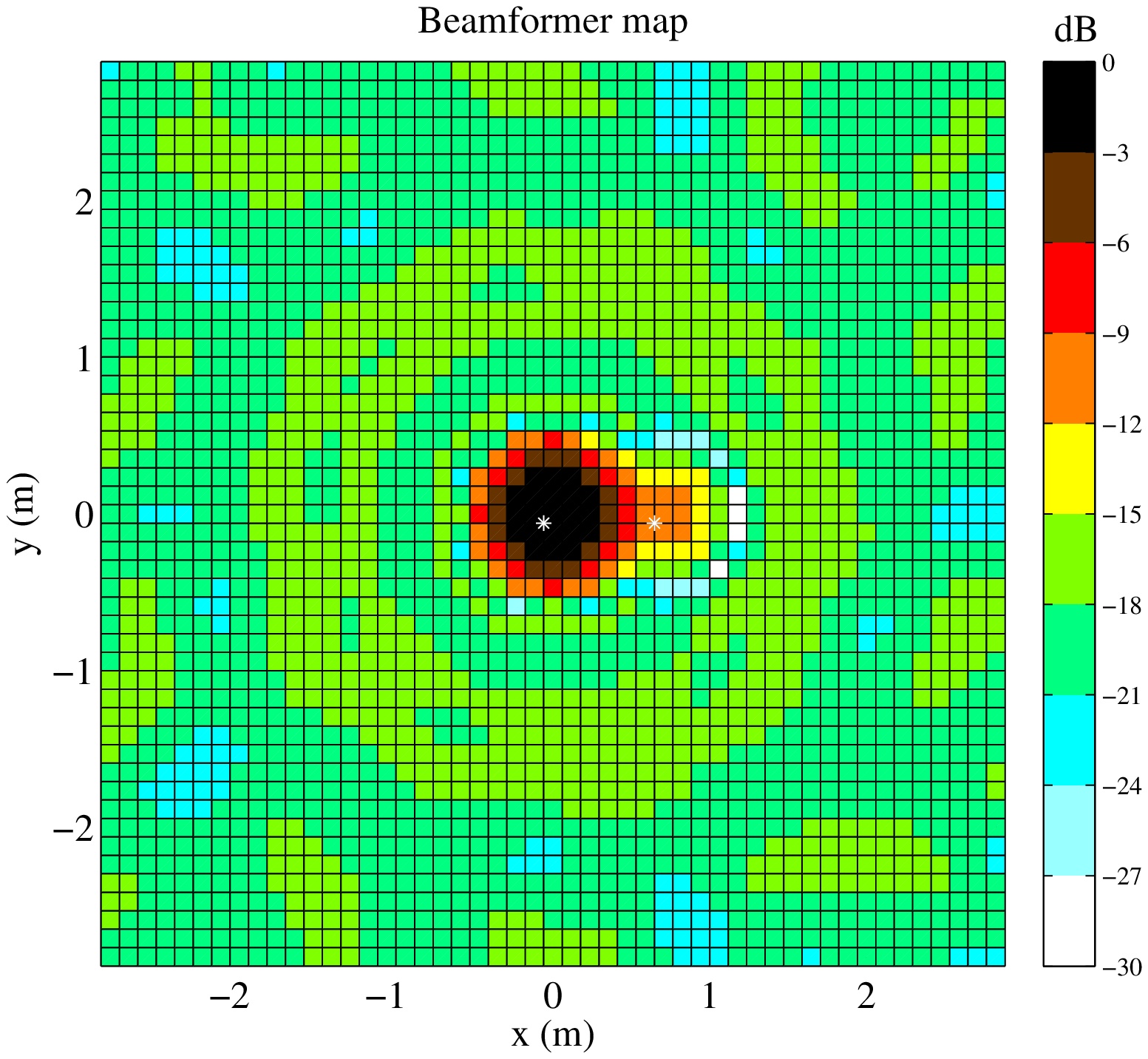}}
  \subfigure[]{
    \label{fig:DAMAS} 
    \includegraphics[width=0.49\textwidth]{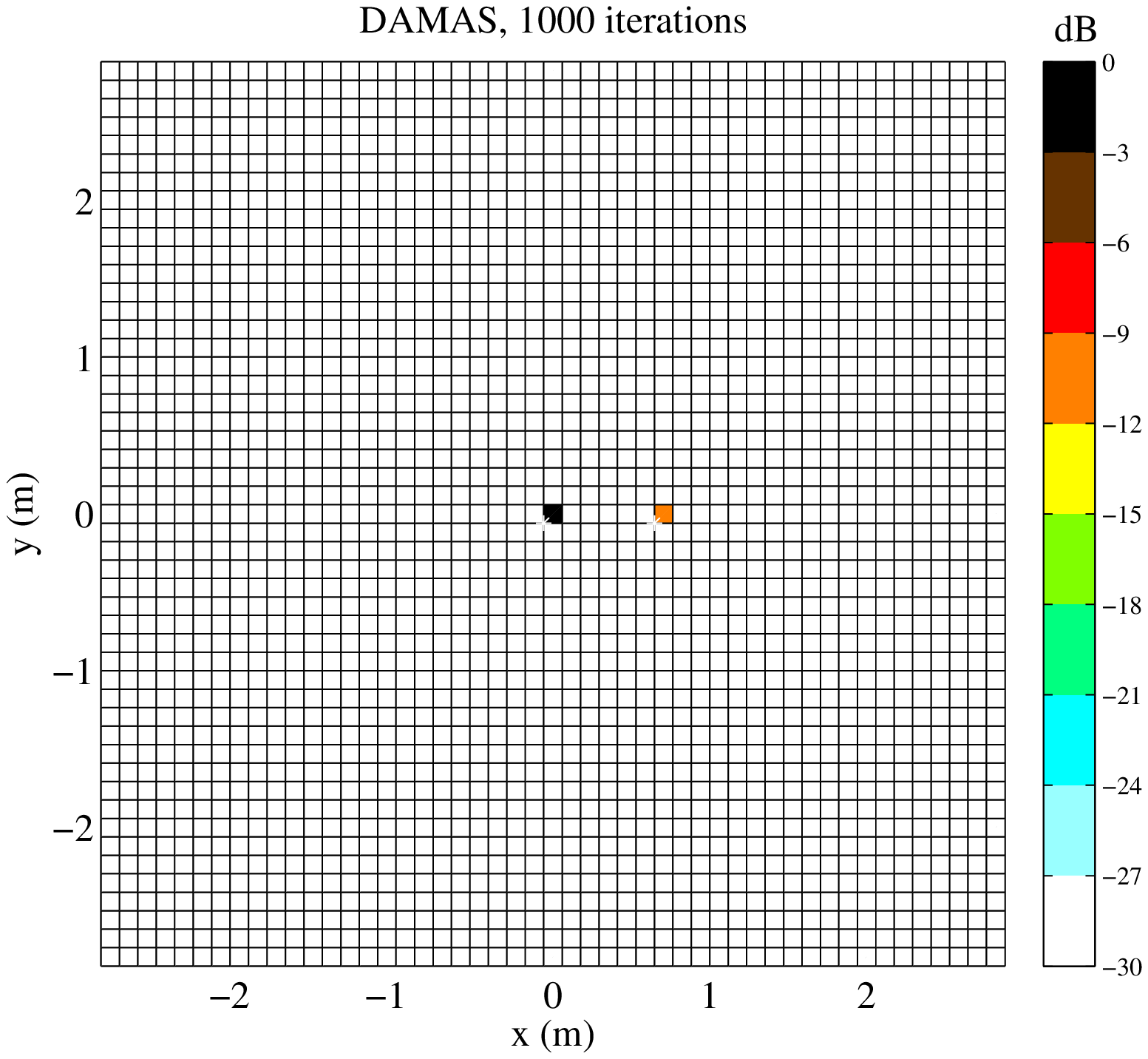}}\\
  \subfigure[]{
    \label{fig:beamformer_map} 
    \includegraphics[width=0.49\textwidth]{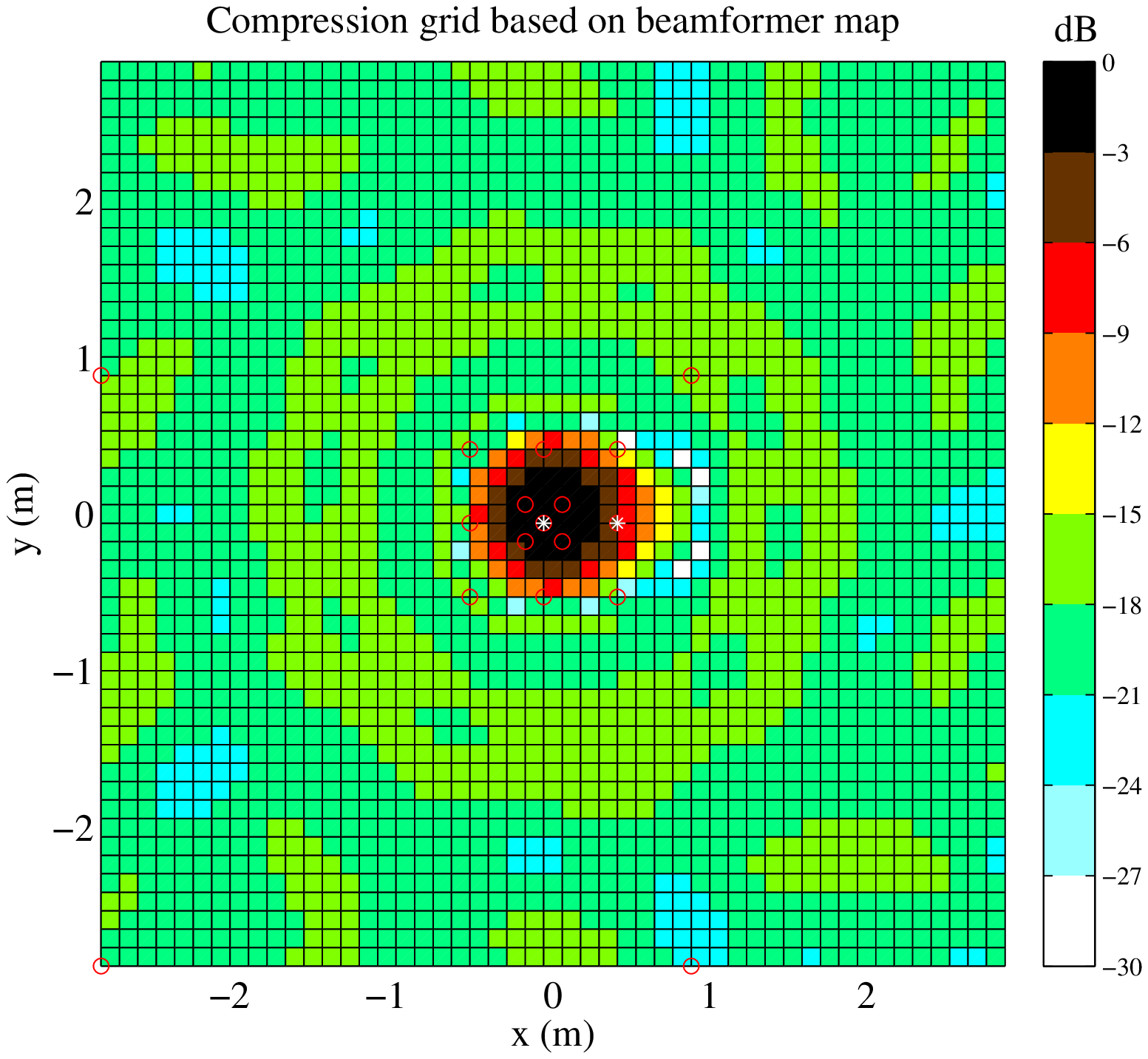}}
  \subfigure[]{
    \label{fig:DAMAS} 
    \includegraphics[width=0.49\textwidth]{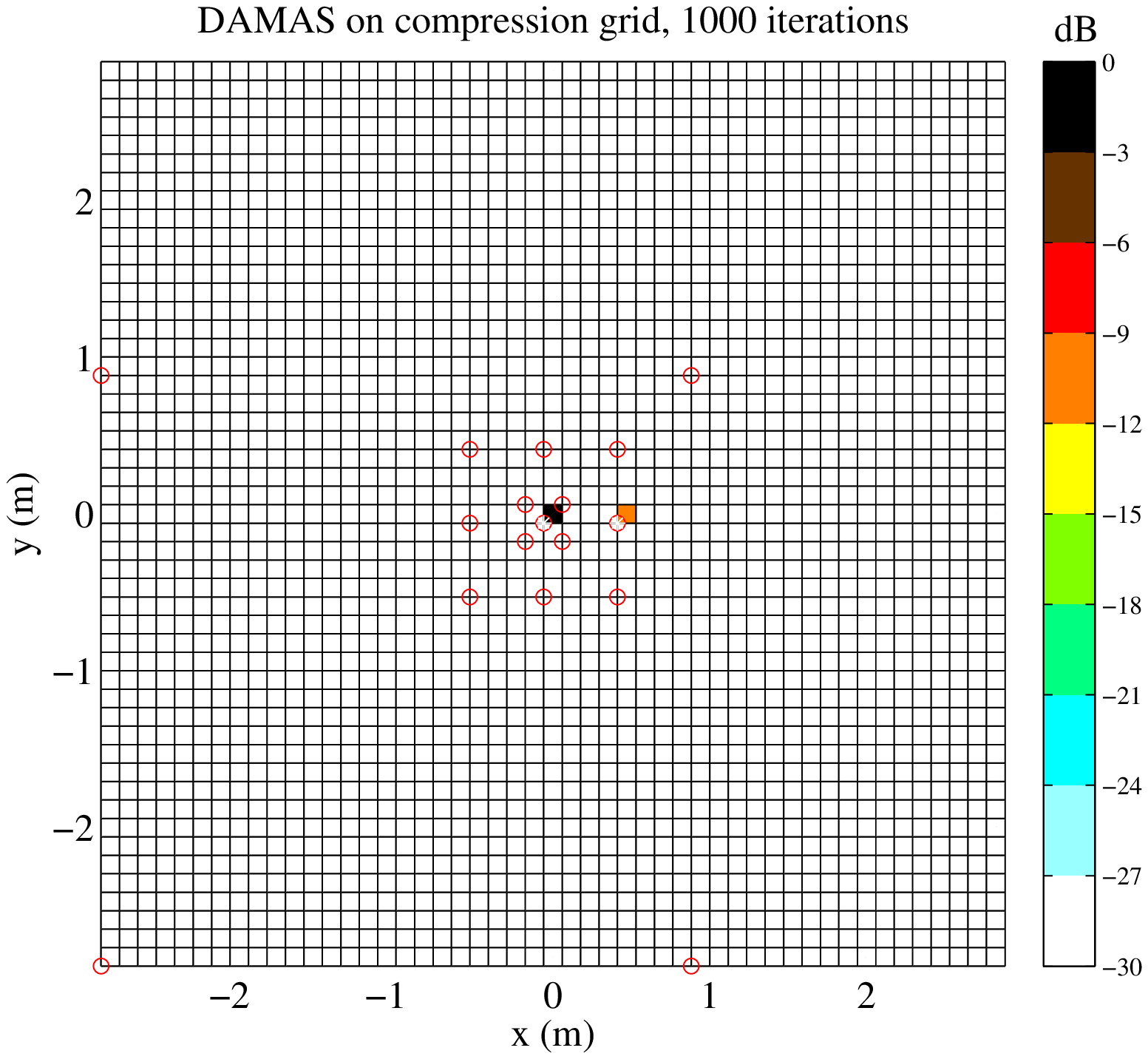}}\\
\caption[]{Synthetic two points source, $f$=3 kHz. White snow symbols, positions of synthetic point sources. (a) Beamformer map on original grid. (b) Deconvolved map of DAMAS on original grid. (c) Compression grid based on beamformer map, shown by red circles. (d) Deconvolved map of DAMAS on compression grid.}
\label{fig:NO3}
\end{figure}

In third simulation, synthetic two point source is placed at grid points (25, 25) and (29, 25).
The source amplitudes $q_s$ are defined to 1.000 Pa at grid point (25, 25), 0.316 Pa at grid point (29, 25), and 0 at all other grid points.
The distance between these two source points is 0.47 m, smaller than beamformer resolution $B$.
The level difference of the two sources is 10 dB.
The corresponding beamformer map is shown in Fig. \ref{fig:NO3}a.
Only the larger source is clearly visible in the beamformer map.
The deconvolved map after applying DAMAS algorithm to the beamformer map is shown in Fig. \ref{fig:NO3}b.
Two sources are well resolved by DAMAS on original grid.

The compression grid based on beamformer map is shown in Fig. \ref{fig:NO3}c with red circle points.
The compression ratio is about 147.
The deconvolved map on the compression grid after applying DAMAS algorithm is shown in Fig. \ref{fig:NO3}d.
Two sources are also well resolved by DAMAS on compression grid.
The error of integrated source power on compression grid is as large as that on original grid.
From the run time listed in Table \ref{tab:appl_simul}, the efficiency of DAMAS at this simulation has also been improved significantly through compression grid as those in the former simulations.

This simulation clearly confirmed that in the dynamic range of 10 dB DAMAS on wavelet compression grid could also improve significantly the efficiency of DAMAS and retain the spatial resolution of DAMAS on original grid.

\subsection{DAMAS image source}
In fourth simulation, DAMAS image source as a complicated source is placed at grid points shown by the snow symbols in Fig. \ref{fig:DAMAS}a.
The source amplitudes $q_s$ are defined to 1.000 Pa at these grid points and 0 at all other grid points.
The corresponding beamformer map is shown in Fig. \ref{fig:DAMAS}a.
In this figure, the structure of source is very difficult to identify.
The deconvolved map after applying DAMAS on the original grid is shown in Fig. \ref{fig:DAMAS}b.
The structure of source is identified at a certain degree.
The error of integrated source power is about 27.0\%.
This means that aliasing already exists in the result of DAMAS on the original grid.

The compression grid based on beamformer map is shown in Fig. \ref{fig:DAMAS}c with red circle points.
The compression ratio is about 4.6.
From the run time listed in Table \ref{tab:appl_simul}, the efficiency of DAMAS in this simulation has been improved significantly through compression grid.
The deconvolved map on the compression grid after applying DAMAS algorithm is shown in Fig. \ref{fig:DAMAS}d.
In this figure, red circles indicating compression grid don't appear for showing more clearly the structure of sound source.
In Fig. \ref{fig:DAMAS}d, the structure of source is substantially similar with that shown in Fig. \ref{fig:DAMAS}b.
The error of integrated source power on compression grid is a little larger than that on original grid.
This implies that aliasing effects increase in the results of DAMAS on the compression grid.
This is reasonable, because the occurrence probability of aliasing increases when the distance between grid points increases in compression grid.

In order to overcome this deficiency, one simple method is changing the density of the grid by adjusting the tolerance $\epsilon$ in Eq. \ref{eq:wavelet_epsilon}.
Fig. \ref{fig:DAMAS}e shows the compression grid with grid points when $\epsilon$=0.05.
The number of grid points at the source region increases obviously compared with that in Fig. \ref{fig:DAMAS}c.
Fig. \ref{fig:DAMAS}f shows the deconvolved map on the compression grid after applying DAMAS algorithm on this compression grid. The structure of source is also identified and similar with that in Fig. \ref{fig:DAMAS}b.
The error of integrated source power is about 27.7\%, smaller than that when $\epsilon$=0.1, and equal to that on original grid.
This means that the aliasing efforts are smaller than that in Fig. \ref{fig:DAMAS}d.
The developments of compression ratio and error of integrated source power with $\epsilon$ are listed in Table \ref{tab:epsilon}.
With the decreasing of $\epsilon$, both compression ratio and error of integrated source power decrease monotonously, meanwhile the run time increases monotonously.
The minimal error of integrated source power is equal to that on original grid.

This simulation verifies that for complicated sound source DAMAS on compression grid largely retains the spatial resolution of DAMAS on original grid, although with a deficiency that the aliasing efforts increase slightly.

\begin{figure}[htbp]
\centering
  \subfigure[]{
    \label{fig:beamformer_map} 
    \includegraphics[width=0.45\textwidth]{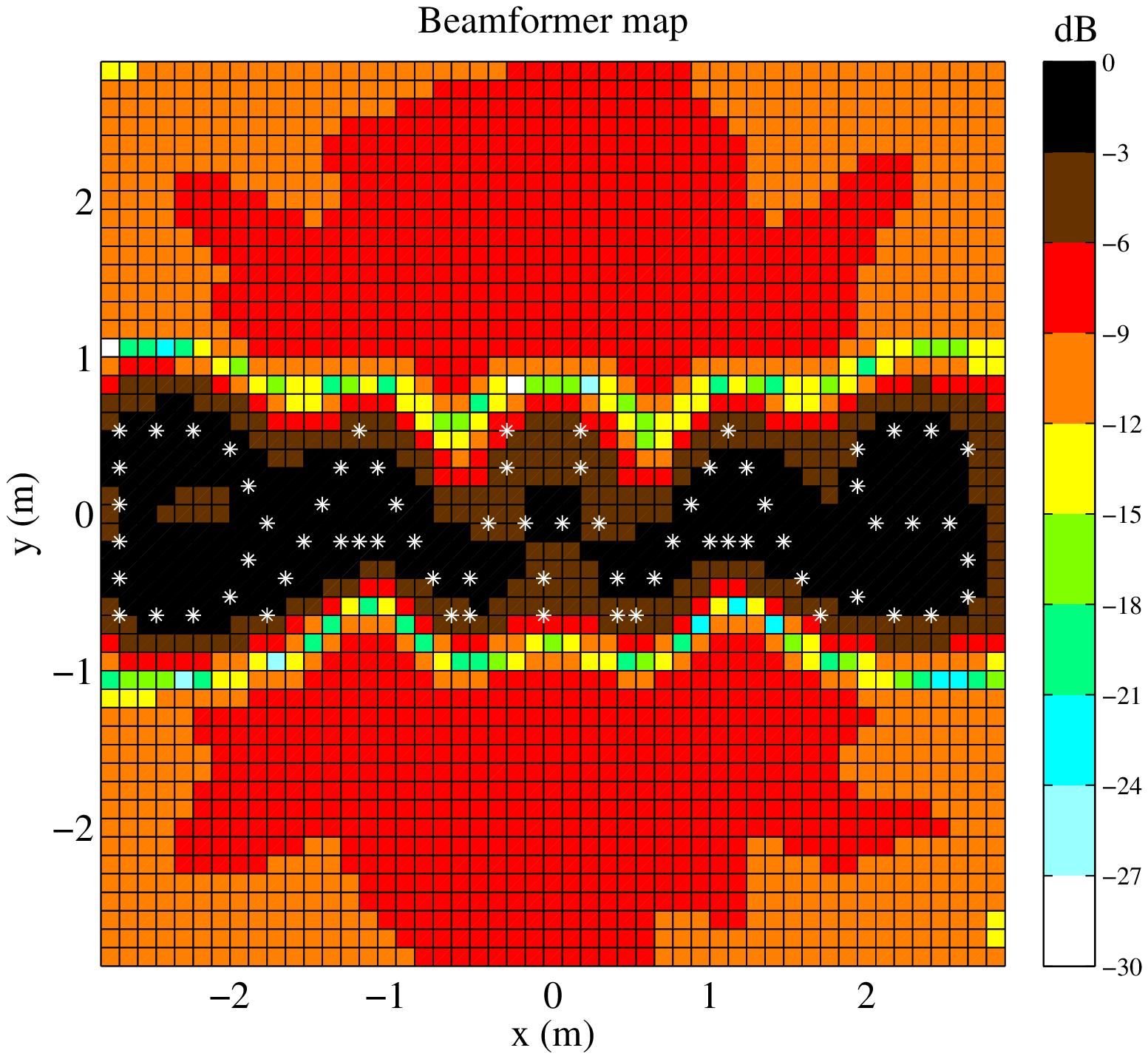}}
  \subfigure[]{
    \label{fig:DAMAS} 
    \includegraphics[width=0.45\textwidth]{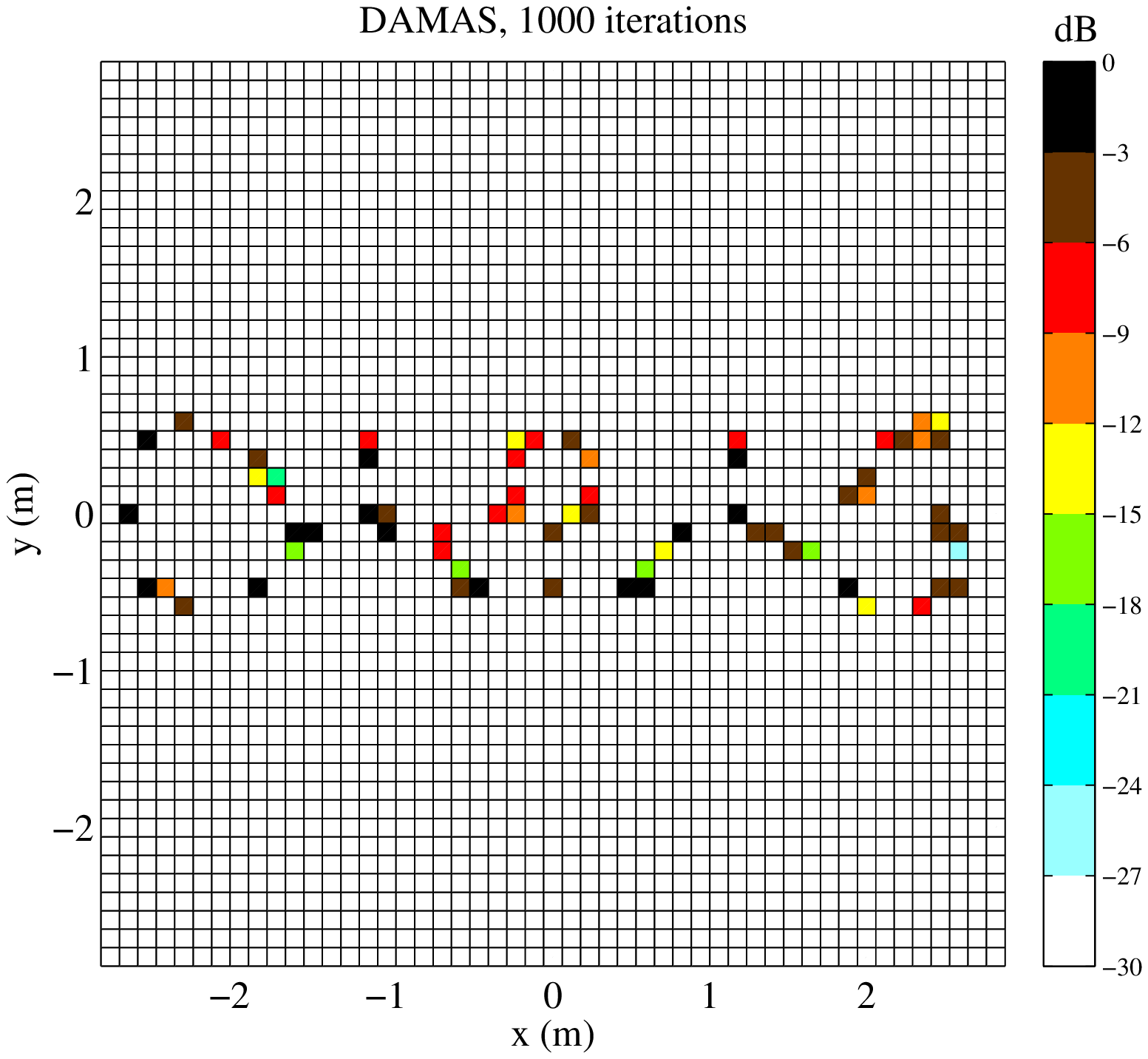}}\\
  \subfigure[]{
    \label{fig:beamformer_map} 
    \includegraphics[width=0.45\textwidth]{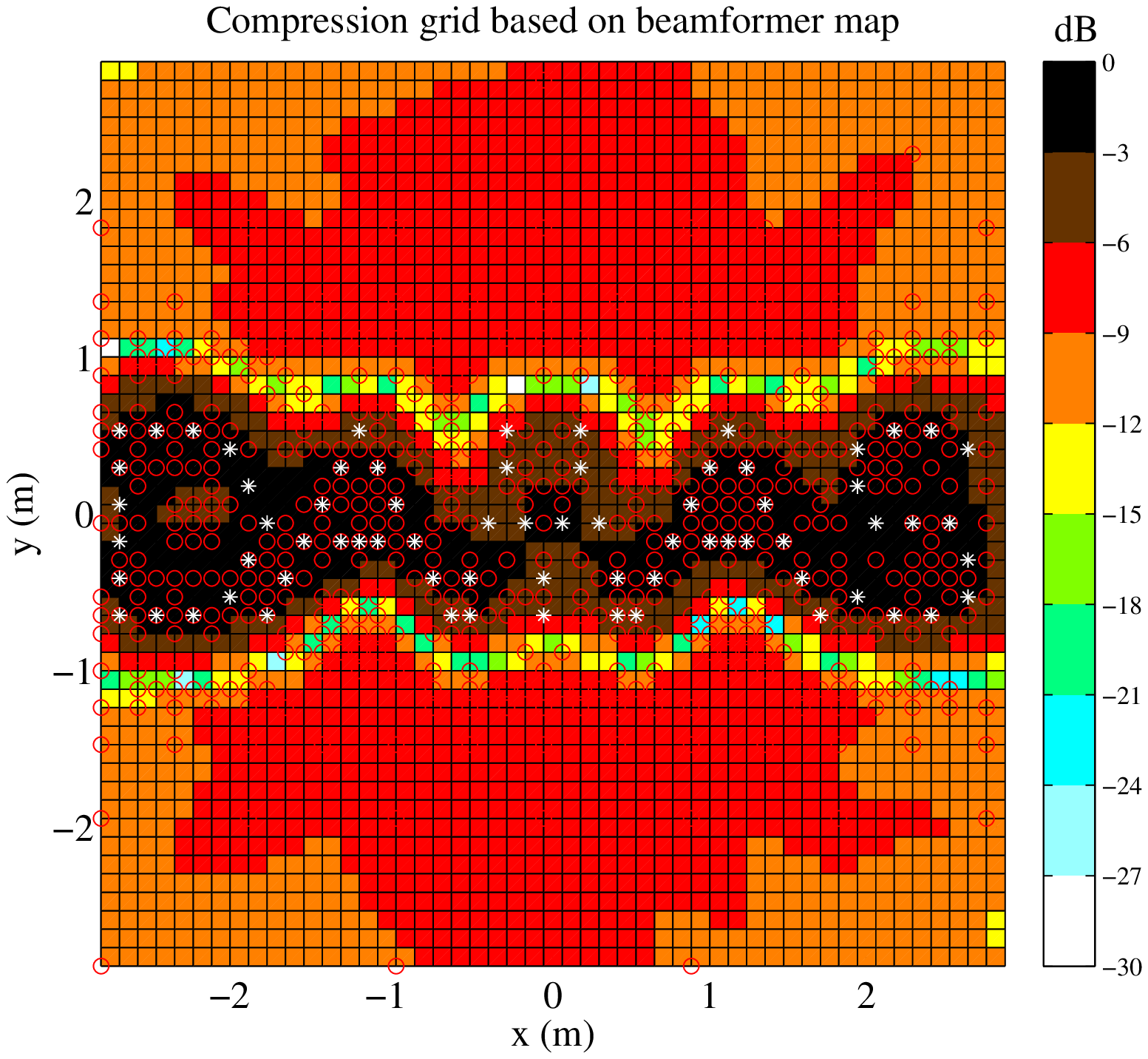}}
  \subfigure[]{
    \label{fig:DAMAS} 
    \includegraphics[width=0.45\textwidth]{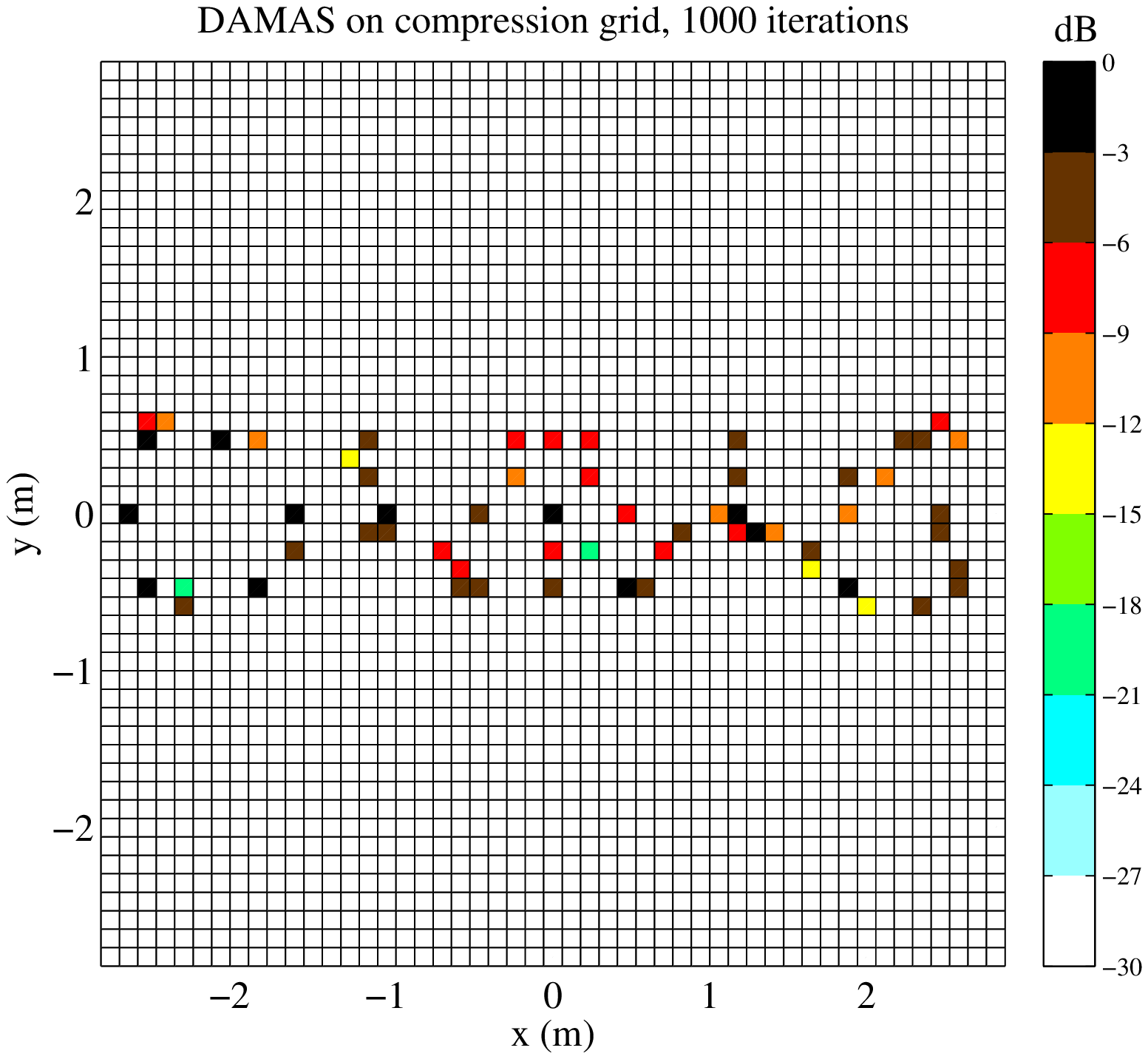}}\\
  \subfigure[]{
    \label{fig:beamformer_map} 
    \includegraphics[width=0.45\textwidth]{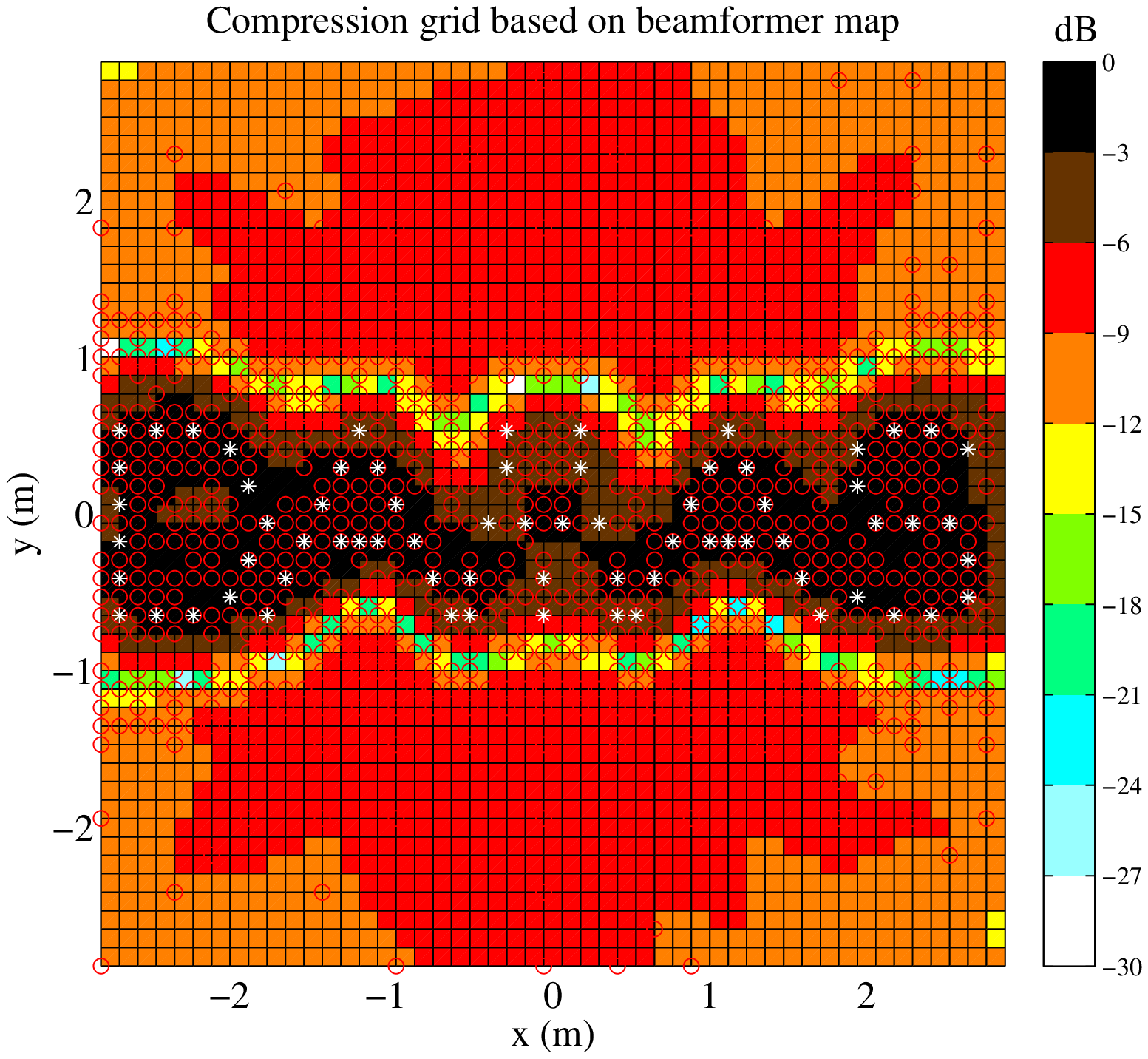}}
  \subfigure[]{
    \label{fig:DAMAS} 
    \includegraphics[width=0.45\textwidth]{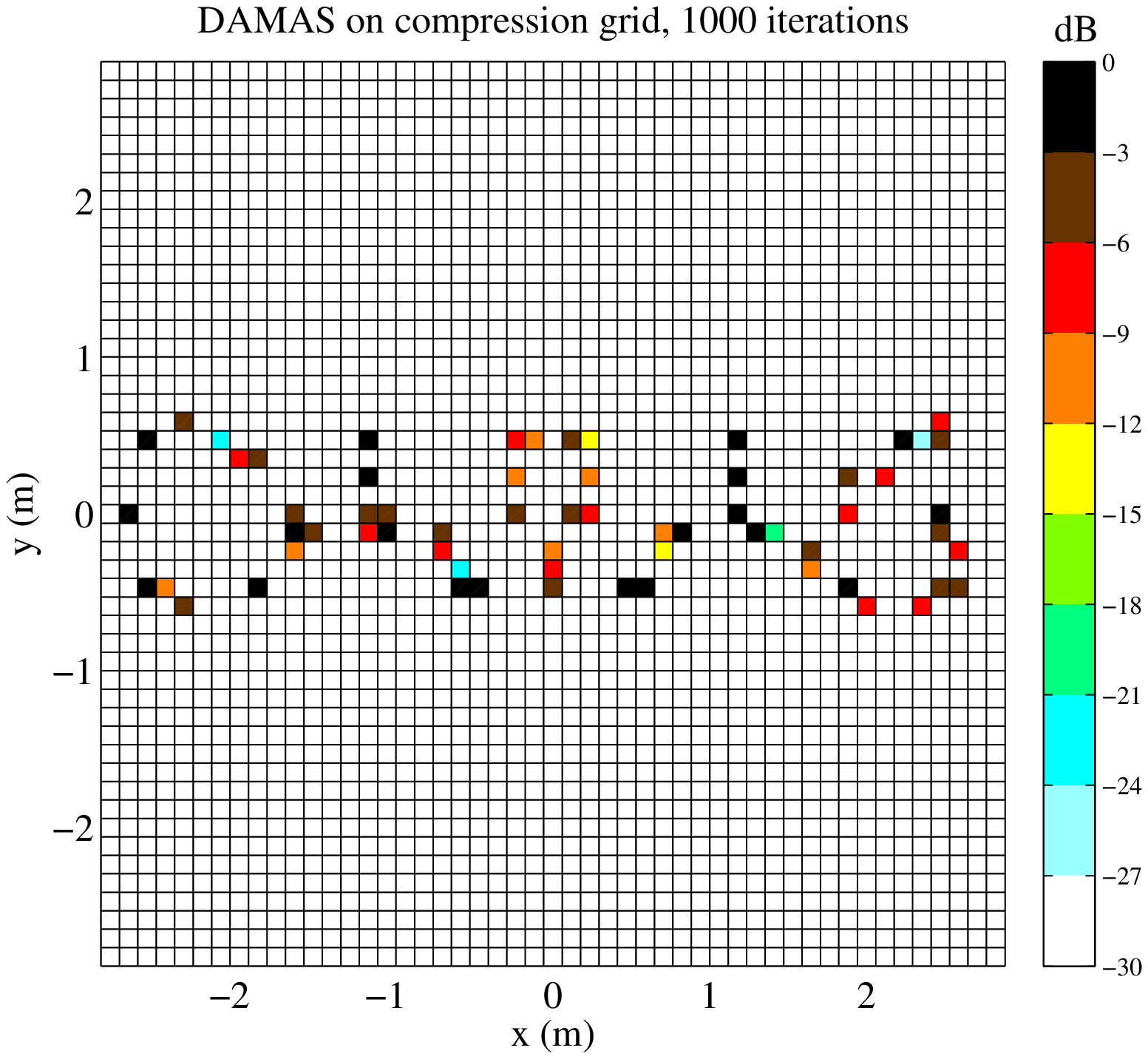}}\\
\caption[]{Synthetic DAMAS image source, $f$=3 kHz. White snow symbols, positions of synthetic point sources. (a) Beamformer map on original grid. (b) Deconvolved map of DAMAS on original grid. (c) Compressed grid based on beamformer map, shown by red circles. $\epsilon$=0.1. (d) Deconvolved map of DAMAS on compression grid shown in c. (e) Compressed grid based on beamformer map, shown by red circles. $\epsilon$=0.05. (f) Deconvolved map of DAMAS on compression grid shown in e.}
\label{fig:DAMAS}
\end{figure}

\begin{table}[htbp]
\centering
\caption{Key application parameter $\epsilon$.}
\label{tab:epsilon}
\begin{tabular}{ccccc}
\hline
 $\epsilon$ &   0.01 & 0.05 & 0.1 & 0.2 \\
 Compression ration, $\sigma$  & 1.8  & 3.1 & 4.6 & 9.7 \\
 Error of integrated source power, $\eta$ & 27.0\%  & 27.7\% & 29.5\% & 31.6\% \\
\hline
\end{tabular}
\end{table}


\clearpage
\section{Discussion}\label{sec:discussion}
In this paper, we propose a new method to improve the efficiency of DAMAS for sound source localization via wavelet compression computational grid.
This is mainly based on three foundational issues.
The first issue is that the computational run time of DAMAS is $O(S^2)$ for a computational grid with $S$ grid points.
The second issue is that wavelet compression according to the beamformer map can identify these significant nodes and discard the remainder redundant points from the original grid for scanning map.
The third issue is that DAMAS could extract substantially similar source distribution on wavelet compression computational grid compared with that on original computational grid.


For run time of this new method, the additional run time is the time for grid compression, while the saved run time is from the iteration time with less total number of grid points.
The time of grid compression usually is even less than the run time of DAMAS on original computational grid for 1 iteration, and thus is negligible compared with the saved run time.
The efficiency of DAMAS has thus been improved significantly via wavelet compression computational grid, and increases with compression ratio.
In industry applications that sound sources are just located in a small extent compared with scanning plane, compression ratio of wavelet compression grid will be much larger than one, and the efficiency of DAMAS could then be improved significantly through wavelet compression computational grid.
In addition, in industrial applications the calculation is usually carried out in a range of angular frequency.
For example, 13 angular frequencies need to be analysed for the range of interest from 400 Hz to 8 kHz using one third octave band filter.
Thus this new method will save a lot of run time.

Concerning spatial resolution, for simple sound source this new method retains the spatial resolution of DAMAS on original grid, even in a dynamic range of 10 dB.
For complicated sound source, this new method could also largely retain the spatial resolution of DAMAS on original grid, however with a minor deficiency that the occurrence probability of aliasing increases.
This deficiency is mainly because the distance between grid points usually increases in compression grid especially in the source region.
This deficiency could be weakened through decreasing the tolerance specified \emph{a priori} in compressing process.
However with a smaller tolerance, the compression ratio will decrease, the run time will increase, and the advantage of this new method becomes weaker subsequently.
For taking advantage and overcoming the deficiency of this new method as much as possible, in industry applications for complicated source, it is recommended to increase appropriately the total number of grid points of the original grid, and chose a appropriate tolerance specified in wavelet compressing process in order to obtain a large compression ration meanwhile make sure as much as possible the $\Delta x/B$ $\leqslant$0.2 in the source region.

In this paper wavelet compression computational grid is used to improve the efficiency of DAMAS for two-dimensional acoustic imaging.
This method is also effective for three-dimensional acoustic imaging \cite{Brooks-2005} which usually demands huge run time.
In addition, wavelet compression computational grid could also be used to improve the efficiency of DAMAS-C \cite{Brooks-2006a} for coherent acoustic sources.
Recently some advanced beamforming algorithms have been proposed to improve the spatial resolution of microphone array such as compressive beamforming \cite{Xenaki-2014, NingFangli-2016} and functional beamforming \cite{Dougherty-2014, Dougherty-2016}.
For improving further the spatial resolution of microphone array, DAMAS could be used to extract source distributions from these new beamformer maps.
Then the efficiency of DAMAS could be improved by using wavelet compression computational grid according to these new beamformer maps.

One of the limiting conditions for spectral procedure applying in deconvolution algorithms is that the number of grid points in each row should be equal.
This limiting condition is not guaranteed to be established in compression computational grid.
As a result, it needs to be investigated further to improve the efficiency of deconvolution algorithms with spectral procedure such as DAMAS2 and FFT-NNLS via wavelet compression computational grid.

\clearpage
\section{Conclusions}\label{sec:conclusions}
In the present paper, the efficiency of DAMAS is improved via wavelet compression computation grid.
To the best knowledge of the authors, this paper is first work so far that address this strategy to improve the efficiency of DAMAS for sound source localization.
In this method, the efficiency of DAMAS increases with compression ratio.
In industrial applications sound source analysis are carried out in a band of angular frequency, thus this method could save lots of run time, particularly when sound sources are just located in a small extent compared with scanning plane.
In addition, DAMAS on wavelet compression computational grid largely retains the spatial resolution of DAMAS on original computational grid, although with a deficiency that the occurrence probability of aliasing increasing slightly for complicated sound source due to the increasing of distance between grid points in compression grid.
This deficiency could be weakened through decreasing the tolerance specified \emph{a priori} in compressing process.
%
Future work should extend the effort to test additional cases, especially where experimental data may be available.

\section*{Acknowledgement}
This work was supported by the NSFC Grants 51506121.

\bibliographystyle{model1-num-names}
\bibliography{../../0-reference-mawei/cornerstall}







\end{document}